\documentclass[showpacs,showkeys,amsmath,amssymb,superscriptaddress,reprint, preprintnumbers,nofootinbib,notitlepage, groupdaddress,showabstract,aps,prd,twocolumn]{revtex4-1}
\pdfoutput=1

\allowdisplaybreaks
\usepackage[pagebackref=false,hidelinks]{hyperref}
\usepackage{setspace,latexsym}
\usepackage{color}
\usepackage{epsfig}
\usepackage{graphicx}
\usepackage{slashed}
\usepackage[export]{adjustbox}
\usepackage{comment}
\usepackage{xcolor}
\usepackage[normalem]{ulem}
\newcommand{\beq}{\begin{equation}}
\newcommand{\eeq}{\end{equation}}
\newcommand{\bea}{\begin{eqnarray}}
\newcommand{\eea}{\end{eqnarray}}
\newcommand{\nn}{\nonumber}

\newcommand{\eV}{\mathrm{eV}}

\newcommand{\GeV}{\mathrm{GeV}}
\newcommand{\MeV}{\mathrm{MeV}}

\newcommand\lsim{\mathrel{\rlap{\lower4pt\hbox{\hskip1pt$\sim$}}
    \raise1pt\hbox{$<$}}}
\newcommand\gsim{\mathrel{\rlap{\lower4pt\hbox{\hskip1pt$\sim$}}
    \raise1pt\hbox{$>$}}}
\def\bal#1\eal{\begin{align}#1\end{align}}

\begin{document}

\title{Cogenesis by a sliding pNGB with symmetry non-restoration}

\author{Eung Jin Chun}
\email{ejchun@kias.re.kr}
\affiliation{Korea Institute for Advanced Study, Seoul 02455, South Korea}

\author{Suruj Jyoti Das}
\email{surujjd@gmail.com}
\affiliation{Particle Theory  and Cosmology Group, Center for Theoretical Physics of the Universe,
Institute for Basic Science (IBS),
 Daejeon, 34126, Korea}

\author{Minxi He}
\email{heminxi@ibs.re.kr}
\affiliation{Particle Theory  and Cosmology Group, Center for Theoretical Physics of the Universe,
Institute for Basic Science (IBS),
 Daejeon, 34126, Korea}

\author{Tae Hyun Jung}
\email{thjung0720@gmail.com}
\affiliation{Particle Theory  and Cosmology Group, Center for Theoretical Physics of the Universe,
Institute for Basic Science (IBS),
 Daejeon, 34126, Korea}

\author{Jin Sun}
\email{jinsun930503@gmail.com}
\affiliation{Particle Theory  and Cosmology Group, Center for Theoretical Physics of the Universe,
Institute for Basic Science (IBS),
 Daejeon, 34126, Korea}

\preprint{CTPU-PTC-24-16}

\begin{abstract}
We demonstrate that a pseudo-Nambu-Goldstone boson (pNGB) with an initial misalignment angle can drive successful spontaneous baryogenesis and serve as a dark matter (DM) candidate, provided the corresponding global symmetry is non-restored at high temperature. A key feature of this mechanism is the presence of a slowly sliding phase in the pNGB’s motion, during which it traverses rapidly diminishing potential barriers, generating and freezing the baryon asymmetry, while transitioning into the kination phase and then an oscillatory phase.
Just before the ‘would-be’ oscillation temperature, parametric resonance effectively fragments the homogeneous mode into fluctuations that ultimately constitute the final DM abundance.
By considering a dimension-five explicit breaking operator, we find that the predicted pNGB mass and decay constant are approximately $5$ eV and $3\times10^6$ GeV, respectively, while the radial mode has a light mass $\mathcal{O}(10)\,{\rm MeV}$ and a small mixing $\mathcal{O}(10^{-4})$ with the Higgs boson. Applied to the Majoron in the type-I seesaw model, this scenario requires the heaviest right-handed neutrino to be as light as $0.1$ to $100\,{\rm GeV}$.
These predictions can be tested through kaon experiments, heavy neutral lepton searches, the LHC, and future colliders.
\end{abstract}

\maketitle

\noindent
\section{Introduction}
The measured matter-antimatter asymmetry 
$n_B/s \simeq 8.75 \times 10^{-11}$\,\cite{Planck:2018vyg}, with $n_B$ and $s$ being the baryon number and entropy densities, is considered as one of the most obvious evidences for new physics beyond the Standard Model (SM).
This asymmetry can be generated dynamically, fulfilling the Sakharov conditions\,\cite{Sakharov:1967dj}, based on which several baryogenesis scenarios are proposed\,\cite{Elor:2022hpa}.

Spontaneous baryogenesis is an attractive scheme where the non-vanishing velocity of a dynamical pseudo-scalar field $a$ (here we consider a pseudo-Nambu-Goldstone boson (pNGB)) induces $CPT$ violation, generating external chemical potentials of quarks and/or leptons\,\cite{Cohen:1987vi, Cohen:1988kt, Domcke:2020kcp}.
The external chemical potentials can be fed thermodynamically into the baryon asymmetry $n_{B}/s\propto \dot \theta/T$ where $\theta\equiv a/f_a$ with the pNGB decay constant $f_a$. 

The crucial part of successful spontaneous baryogenesis is to generate sufficiently large $\dot \theta$ in the early Universe. 
For instance, $\dot{\theta}$ can be generated in the conventional misalignment mechanism as originally discussed in Refs.\,\cite{Cohen:1987vi, Cohen:1988kt} although it oscillates and requires some tuning of different time scales.
In the kinetic misalignment (KM) mechanism\,\cite{Co:2019jts, Co:2019wyp}, $\dot \theta$ is generated dynamically from an initial large radial field value with an explicit symmetry-breaking operator, analogously to the Affleck-Dine mechanism\,\cite{Affleck:1984fy}, which has been applied to various models\,\cite{Co:2019jts, Co:2019wyp, Co:2020xlh, Co:2020jtv, Harigaya:2021txz, Chakraborty:2021fkp, Kawamura:2021xpu, Co:2021qgl, Co:2022aav, Barnes:2022ren, Co:2022kul, Chun:2023eqc, Badziak:2023fsc, Berbig:2023uzs, Chao:2023ojl, Chang:2024xjd, Barnes:2024jap, Datta:2024xhg}.
An alternative is first-order phase transition where $\partial_\mu \theta$ is generated across bubbles\,\cite{Jeong:2018ucz, Jeong:2018jqe, Harigaya:2023bmp}.

In this paper, we propose a new mechanism for generating $\dot \theta$.
We consider an initial misalignment angle 
$ \theta_i $ while we assume that the radial mode follows its potential minimum point adiabatically, which seems similar to the initial setup of the conventional misalignment mechanism.
The difference comes from the framework of the symmetry non-restoration\,\cite{Weinberg:1974hy, Mohapatra:1979qt, Fujimoto:1984hr, Salomonson:1984rh, Bimonte:1995sc, Dvali:1995cj, Dvali:1996zr, Orloff:1996yn, Gavela:1998ux, Bajc:1998jr, Espinosa:2004pn, Ahriche:2010kh, Kilic:2015joa, Meade:2018saz, Baldes:2018nel, Glioti:2018roy, Matsedonskyi:2020mlz, Matsedonskyi:2020kuy, Agrawal:2021alq, Biekotter:2021ysx, Carena:2021onl, Matsedonskyi:2021hti, Chang:2022psj}, where the decay constant scales as $f_a \propto T$ at high temperature.
This induces two effects: (i) it reduces the Hubble friction of $\theta$ dynamics, and (ii) the pNGB potential barrier decreases quickly as temperature drops.
Consequently, there appears a long period of \emph{pNGB sliding} during which $\dot \theta/T$ becomes sizable and approximately constant 
until the onset of oscillation.
It can result in successful baryogenesis during sliding, while the DM abundance is related to the zero-temperature values of $m_a$ and $f_a$, preventing overproduction.

Our scenario has a close resemblance with that of KM\,\cite{Co:2019jts}, where the pNGB develops enough kinetic energy to cross the potential barrier, gets trapped later, and oscillates.
However, unlike KM, we do not require the radial mode of the symmetry-breaking field to be initially far away from the potential minimum. 
In our case, its value follows the minimum of the thermal effective potential.
Subsequently, we attain  $\dot \theta\propto T$ (in contrast to $\dot \theta\propto T^3$ in KM) while sliding across the decreasing potential barriers.

Another similar dynamics can be found in so-called axion monodromy\,\cite{Kim:2004rp,McAllister:2008hb, Silverstein:2008sg}, where multiple periodicities appear in the potential, which have been studied in various contexts such as inflation\,\cite{Kim:2004rp, McAllister:2008hb, Silverstein:2008sg, Pajer:2013fsa, Marchesano:2014mla}, dark matter\,\cite{Jaeckel:2016qjp} or relaxion\,\cite{Graham:2015cka}.
The origin of such multi-periodic potentials typically requires hierarchically different symmetry-breaking scales\,\cite{Kim:2004rp}, unlike the single scale $f_a$ in our case.
In spite of the completely different setup, the solution of our pNGB dynamics is similar to the axion monodromy, and thus we apply the fragmentation analysis well-studied in the axion monodromy\,\cite{Jaeckel:2016qjp,Berges:2019dgr, Fonseca:2019ypl} and also the KM context\,\cite{Eroncel:2022vjg, Eroncel:2022efc} to our case.

\section{Global symmetry non-restoration}
Consider a complex scalar field $\Phi$ with zero-temperature potential 
\bal
V(\Phi) =
\lambda_\phi |\Phi|^4 -m_0^2 |\Phi|^2.
\label{Eq:Vphi}
\eal
Its vacuum expectation value (vev) $\langle |\Phi| \rangle = m_0/\sqrt{2\lambda_\phi}\equiv f_a^{(0)}/\sqrt{2}$
breaks the global $U(1)$ symmetry spontaneously, resulting in a NG mode $a$ with decay constant $f_a$ in the decomposition $\Phi = \frac{1}{\sqrt{2}}\phi \, e^{ia/f_a}$.

The mass of $a$ arises from explicit $U(1)$ symmetry breaking via higher-dimensional operators
\bal
\frac{\Phi^n}{\Lambda^{n-4}} + {\rm H.c.}
\Rightarrow
V_a(a) = \frac{f_a^n}{2^{\frac{n}{2}-1} \Lambda^{n-4}} \! \left( \! 1-\cos \! \left( \! \frac{n a}{f_a} \! \right) \!\right)\! ,\!
\label{Eq:Va}
\eal
where the integer $n>4$, and $\Lambda$ is the  cutoff scale of the dimension $n$ operator, treated as a free parameter traded with $m_a$. Here, we shifted $a/f_a$ by a constant to make $V_a(a)$ minimized at $a=0$ and added a constant to make $V_a(0)=0$

At high temperatures, the potential \eqref{Eq:Vphi} receives thermal corrections.
In particular,  a negative $T^2$ mass correction to $\Phi$ can arise, which prohibits symmetry restoration. 
This can be realized when $\Phi$ couples to the SM Higgs doublet $\mathcal{H}$ via  $\Delta {V} = -2\lambda_{h\phi} |\mathcal{H}|^2|\Phi|^2$ with $\lambda_{h\phi} >0$ in the minimal setup, while a general extension introduces $N_s$ real scalar fields $s_i$  with 
$\Delta {V} = -\lambda_{\phi s_i} |\Phi|^2 s_i^2 $ and $ \lambda_{\phi s_i}>0 $.
In either case, we will use $\lambda_{\rm mix}\equiv \lambda_{h\phi} + \sum_i \lambda_{\phi s_i}/4$ to denote such quartic couplings. 
Meanwhile, we ensure that the electroweak symmetry gets restored at high temperatures due to gauge and Yukawa couplings.

Then, the thermal potential for $\phi$ can be written as
\bal
V_T(\phi) \simeq
\frac{\lambda_\phi}{4} \phi^4 -\frac{1}{2}(m_0^2 + c \, T^2) \phi^2,
\label{Eq:VTphi}
\eal
with a model-dependent constant $c$.
This potential is minimized at
\bal
&\langle \phi \,\rangle_T = f_a(T) = \sqrt{{f_a^{(0)}}^2+ c_\lambda T^2},\label{eq:faT}
\eal
with $c_\lambda \equiv c/\lambda_\phi$ and the pNGB mass becomes
\bal
&m_a^2(T) =\frac{n^2}{2^{\frac{n}{2}-1} }
\left( \frac{f_a(T)}{\Lambda}\right)^{n-4}
f_a(T)^2.
\label{Eq:maT}
\eal
Thus, at $T>T_c \equiv f_a^{(0)}/\sqrt{c_\lambda}$, $f_a(T)$ and $m_a(T)$ scale as $T$ and $T^{(n-2)/2}$, respectively.
We obtain $c \simeq \lambda_{\rm mix}/3$ and $c_\lambda \simeq \lambda_{\rm mix}/3\lambda_\phi$.

The parameter
$c_\lambda$ has to be large enough to avoid affecting the matter-radiation equality by pNGB production through the freeze-in process $\phi\phi \to a a$.
The averaged cross-section of $\phi\phi \to a a$ in thermal equilibrium is roughly $T^2/f_a^4\sim 1/(c_\lambda^2  T^2)$ for $T>T_c$ and gets suppressed when $T<T_c$.
For $m_a >\eV$, its relic abundance must be smaller than that of DM, and we obtain the lower bound of $c_\lambda \gsim 10^{7}$ which can be realized by taking $\lambda_{\rm mix}\lsim 10^{-8}$ with $\lambda_\phi \sim \lambda_{\rm mix}^2$.
In Appendix \ref{ApA}, we show that these small couplings are not messed up by quantum corrections.

$\phi$ necessarily mixes with the SM Higgs  $h$ to evade stringent Big Bang nucleosynthesis (BBN) constraints\,\cite{Fradette:2017sdd,Goudzovski:2022vbt}, even when negative $T^2$ corrections arise dominantly from other fields. With mass \mbox{$m_{\phi}  \simeq \sqrt{2\lambda_{\phi}} f_a^{(0)} \ll m_h$} (the scale of $f_a^{(0)}\sim 10^6\,\GeV$ will be given later), the mixing angle becomes $\sin \theta_{h \phi} \simeq  2 \lambda_{h\phi} v_h f_a^{(0)}/m_h^2$ with the Higgs vev $v_h=246\,\GeV$ and the Higgs mass $m_h = 125\,\GeV$.
To make $\phi$ decay before BBN, $\phi$ has to be heavier than twice the electron mass, with $\sin \theta_{h\phi}\gtrsim 10^{-5}$.
Then, the upper bound of $\sin \theta_{h\phi} \lsim 10^{-4}$ is given from kaon experiments\,\cite{Choi:2016luu, Flacke:2016szy, Beacham:2019nyx, Agrawal:2021dbo,  Goudzovski:2022vbt,NA62:2021zjw,NA62:2020pwi,NA62:2020xlg,BNL-E949:2009dza,KOTO:2018dsc,KOTO:2020prk,Kitahara:2019lws}, for $m_\phi\lesssim 100\,\MeV $.

\section{Dynamics before oscillation}
Let us now discuss the dynamics of the pNGB angle $\theta \equiv a/f_a(T)$ when $ T \gsim T_c $.
With the homogeneity of $\theta$, its equation of motion is 
\bal
\ddot \theta + \left(3H +2\frac{\dot f_a}{f_a} \right)\dot \theta = - \frac{1}{n} m_a^2(T) \sin (n \theta),
\label{Eq:eom}
\eal
where $\phi$ is assumed to follow its potential minimum adiabatically which will be examined later with an explicit model.

In the following, we focus on $n=5$ because the scenario is independent of the non-adiabatic evolution of $\phi$ in the initial stage when $\phi$'s effective potential is developed.
Later, we will give an example UV model where the dimension-five explicit breaking operator dominates over other explicit breaking operators, and also provide a qualitative sketch of how the scenario changes for $n\neq 5$.

For $n=5$, from Eq.\,\eqref{Eq:maT} we have $m_a(T) \simeq 3\,c_\lambda^{3/4}(T/\Lambda)^{3/2}\Lambda \propto T^{3/2}$.
At high temperature when  $H>m_a(T)$, $\theta$ is nearly frozen at $\theta_i$.
Using $\dot f_a/f_a \simeq -H$, we find $\dot \theta/T \propto \log (T/T_{\rm RH})$, where $T_{\rm RH}$ is the reheating temperature.

The $\log T$ scaling behavior lasts until $H(T)$ becomes comparable to the pNGB mass $m_a(T)$.
Thus, it is convenient to define this characteristic temperature $T_0$ when $H(T_0)=m_a(T_0)$;
\bal
T_0\simeq 5 \!\times \! 10^{11}\,\GeV
\bigg( \! \frac{100}{g_*} \! \bigg) 
\bigg( \! \frac{c_\lambda}{10^8} \! \bigg)^{\!\! \frac{3}{2}} \!
\bigg( \! \frac{m_a^{(0)}}{\eV} \! \bigg)^{\!\! 2} \!
\bigg( \! \frac{10^6\,\GeV}{f_a^{(0)}} \! \bigg)^{\!\! 3}
\label{eq:T0}
\eal
where we used $m_a(T_0) \simeq m_a^{(0)} (T_0/T_c)^{3/2}$.
\begin{figure}
    \centering
    \includegraphics[width=0.48\textwidth]{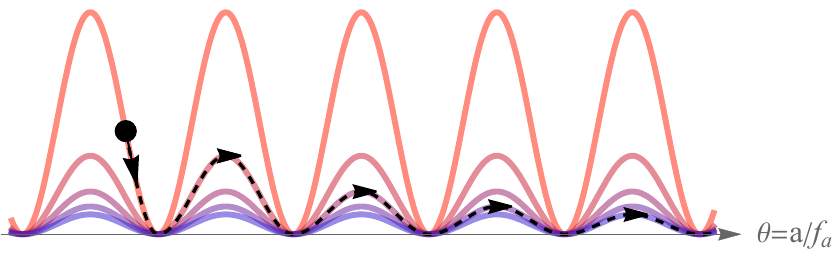}
    \caption{Schematic plot of the pNGB slide.}
    \label{Fig:schematic}
\end{figure}

With $\theta_i$ not too small, the pNGB starts sliding over the potential below $T_0$, instead of oscillating (see Fig.\,\ref{Fig:schematic}).
This is because the potential height of $\theta$ given by $m_a^2\propto T^3$  decreases faster than the redshift of the kinetic energy, $\dot \theta^2 \propto T^2$.
During this period, $\dot \theta/T$  is approximately constant, since  Eq.\,\eqref{Eq:eom} can be approximated as $\ddot \theta + H \dot \theta =  T \frac{d}{dt}(\dot \theta/T)=0$.
However, if $ \theta_i $ is too small, then the pNGB may not slide as soon as $T$ approaches $T_0$ since the initial velocity $\dot \theta$ is too small to climb up the potential.
Then, it first oscillates a few periods before sliding, with $\theta \simeq A(t) \cos[\int m_a\, dt]$, where the amplitude $A$ follows $A^2 m_a \propto T$. 
During oscillation, $\dot \theta^2 \simeq m_a^2 A^2$ scales as $T^{5/2}$, so the kinetic energy can overcome the potential at some point and the pNGB starts sliding away.
We estimate this sliding temperature as
\bal
T_{\rm slide} \simeq
\frac{\mathcal{C}}{4}
T_0
(1-\cos(5\theta_i))^2,
\label{eq:Tslide}
\eal
with a numerical factor ${\cal C} \sim \mathcal{O}(10)$ which depends on $T_{\rm RH}$ logarithmically with some power.
In either case, $\dot \theta/T$ during the slide is constant and maximal in the history of the pNGB dynamics.
Its value is given by
\bal
&\bigg(\frac{\dot \theta}{T}\bigg)_{\rm \!\! slide} \!\!\!\!\!\!\!
\simeq
\frac{2m_a(T_{\rm slide})}{5T_{\rm slide}}
\label{Eq:thdotmax}
\\
&  \simeq
7 \!\times\! 10^{-8} \, {\cal C}^{\frac{1}{2}} (5\theta_i)^2
\bigg( \! \frac{100}{g_*} \! \bigg)^{\!\! \frac{1}{2}} \!
\bigg( \! \frac{c_\lambda}{10^8} \! \bigg)^{\!\! \frac{3}{2}} \!
\bigg( \! \frac{m_a^{(0)}}{\eV} \! \bigg)^{\!\! 2} \!
\bigg( \! \frac{10^6\,\GeV}{f_a^{(0)}} \! \bigg)^{\!\! 3}.
\nn
\eal

\section{Oscillation energy density}
For $T<T_c$, $f_a$ and $m_a$ get saturated to the zero-temperature values so $\dot f_a/f_a =0$ leading to $\dot \theta\propto T^{3}$ during sliding motion, like the usual kination. The final oscillation starts when $\dot \theta(T_{\rm osc}) \simeq \frac{2}{5} m_a^{(0)}$ from which we obtain
\bal
\label{Eq:Tosc}
T_{\rm osc}
&\simeq
\frac{4\,\GeV}{{\cal C}^{\frac{1}{6}} (5\theta_i)^{\frac{2}{3}}}
\bigg( \! \frac{g_*}{100} \! \bigg)^{\!\! \frac{1}{6}} \!
\bigg( \! \frac{10^8}{c_\lambda} \! \bigg)^{\!\! \frac{5}{6}} \!
\bigg( \! \frac{\eV}{m_a^{(0)}} \! \bigg)^{\!\! \frac{1}{3}} \!
\bigg( \! \frac{f_a^{(0)}}{10^6\,\GeV} \! \bigg)^{\!\! \frac{5}{3}}  .
\eal
This is smaller than the typical oscillation temperature in the standard misalignment mechanism. 

The ratio of the oscillation energy density and the entropy density is frozen as
\bal
&\frac{\rho_{\rm osc}}{s}
\simeq
2\times\frac{\frac{2}{25} \big({m_a^{(0)}} {f_a^{(0)}}\big)^2}{s(T_{\rm osc})}
\label{Eq:DM}
\\
&\simeq
0.07\,\eV \,
{\cal C}^{\frac{1}{2}} {(5\theta_i)^{2}}
\bigg( \! \frac{100}{g_*} \! \bigg)^{\!\! \frac{3}{2}} \!
\bigg( \! \frac{c_\lambda}{10^8} \! \bigg)^{\!\! \frac{5}{2}} \!
\bigg( \! \frac{m_a^{(0)}}{\eV} \! \bigg)^{\!\! 3} \!
\bigg( \! \frac{10^6\,\GeV}{f_a^{(0)}} \! \bigg)^{\!\! 3}  ,
\nn
\eal
where the factor $\simeq 2$ comes from our numerical checks (see Appendix \ref{ApB} for details).
Requiring  $\rho_{\rm osc}/s = 0.44\,\eV$\, for the observed DM abundance\,\cite{Planck:2018vyg}, we obtain the relation between $m_a^{(0)}$ and $f_a^{(0)}$ for a given $c_\lambda$. 

Note that the above estimate and the scenario of sliding across the potential barriers, and finally getting trapped, remains essentially the same as KM\,\cite{Co:2019jts}.
However, in our case,  the oscillation abundance is determined by the initial misalignment angle $\theta_i$ and $f_a^{(0)}, m_a^{(0)}, c_{\lambda}$, which set the initial $\dot{\theta}$ (cf. Eq.\,\eqref{Eq:thdotmax}). This is different from KM, where $\dot \theta$ is determined crucially by a large initial value of $\phi$.

\section{\lowercase{p}NGB fragmentation}
Our estimation so far assumed a homogeneous $\theta$, but fluctuations of $\theta$ can be enhanced via parametric resonance\,\cite{Berges:2019dgr,Fonseca:2019ypl, Eroncel:2022vjg}. A fluctuation mode $k$ is resonantly enhanced when the modulation frequency $\dot\theta$ matches roughly $k/R$, where $R$ is the scale factor. How precisely the resonance condition needs to be satisfied can be quantified by the height of the potential barrier, which can be roughly written as 
\bal
|(k/R)^2- \dot \theta^2| \lsim m_a^2,
\eal
called the resonance band (see Appendix \ref{ApC} for more precise expressions). Modes staying in the resonance band for an extended period grow exponentially, draining energy from the homogeneous mode and causing its fragmentation. If that happens, the sliding stops at the fragmentation time (before $T_{\rm osc}$), and the final DM abundance must be modified properly.

At $T>T_c$, we have $\dot \theta \propto T$, causing some specific $k$ modes to be inside the resonance band for a long time because $R\propto T^{-1}$, while the band width shrinks. This imposes a constraint of $5\theta_i \gtrsim 0.7$ to prevent fragmentation, as determined numerically (see Appendix \ref{ApC}). For simplicity, we focus on $5\theta_i \gtrsim 0.7$ hereafter, although we emphasize that avoiding fragmentation at $T>T_c$ is not strictly necessary, as baryogenesis could conclude before fragmentation.

Once pNGB sliding is maintained until $T_c$, the dynamics at $T<T_c$ is exactly the same as the KM scenario\,\cite{Eroncel:2022vjg}, where it was shown that the growth of the fluctuations due to parametric resonance depends crucially on the ratio $m_a^{(0)}/H(T)$ at the time of trapping. Since this ratio turns out to be quite large in our scenario, we find that complete fragmentation occurs slightly before $T_{\rm osc}$ in Eq.\,\eqref{Eq:Tosc}, which should be now called \emph{would-be} oscillation temperature. Thus, DM is not a coherent oscillation, but the pNGB fluctuations.
However, the final DM abundance matches the estimation of the unfragmented scenario, which can be understood as follows (shown explicitly in Appendix \ref{ApC}). 
Since the fragmentation takes place slightly before $T_{\rm osc}$, a higher energy density is assigned to the DM component (the pNGB fluctuation).
However, the enhanced fluctuations are slightly relativistic when they are generated, and hence their energy density dilutes more rapidly than the coherent oscillation until they become non-relativistic. 
The redshift factors from these two contributions scale exactly in the opposite way, and cancel each other up to $\mathcal{O}(1)$ factors, leading to the same DM abundance as in Eq.\,\eqref{Eq:DM}.

\section{An explicit model}
To discuss how successful baryogenesis arises, let us consider the type-I seesaw model extended with the Majoron as an example.
The model Lagrangian includes
\bal
-\Delta {\cal L} = \left(\sum_{i} y_{i} \Phi \bar{N}_i^c N_i + \sum_{i,\alpha} Y_{D, {\alpha i}}\bar{l}_\alpha \Tilde{\mathcal{H}} N_i +h.c.\right) + V(\Phi)\,,
\eal
where $N_i$ and $l_\alpha$ denote the right-handed neutrino (RHN) and the SM lepton doublet, respectively, with three generations each. 
Here, the anomaly-free $U(1)_{B-L}$ global symmetry is imposed by hand with $\Phi$ and $N_i$ having $B-L$ charges $2$ and $-1$, while the vev of $\Phi$ breaks it spontaneously.
After this symmetry breaking, the model fits into the conventional type-I see-saw model\,\cite{Minkowski:1977sc,Yanagida:1980xy,Gell-Mann:1979vob,Mohapatra:1979ia,Schechter:1980gr} with the Majorana masses of $N_i$ being $M_{N_i}=y_i f_a^{(0)}/\sqrt{2}$, explaining small neutrino masses, given as \mbox{$(m_\nu)_{\alpha \beta} \sim \sum_i Y_{D, {\alpha i}}Y_{D, {\beta i}}^* v_h^2/2M_{N_i}$}.
In this setup, we apply the previous discussion to the $\Phi$ sector, and its axial mode, the Majoron, plays the role of the pNGB. 
We require $y_i \sqrt{c_\lambda} \lsim 4\pi$ and $y_i^2 < 4\lambda_{\rm mix}$ to avoid spoiling the potential of $\phi$ with small $\lambda_\phi$ since Yukawa interactions generate quantum and thermal corrections. 

\begin{figure}[t]
    \centering
   \includegraphics[width=0.48\textwidth]{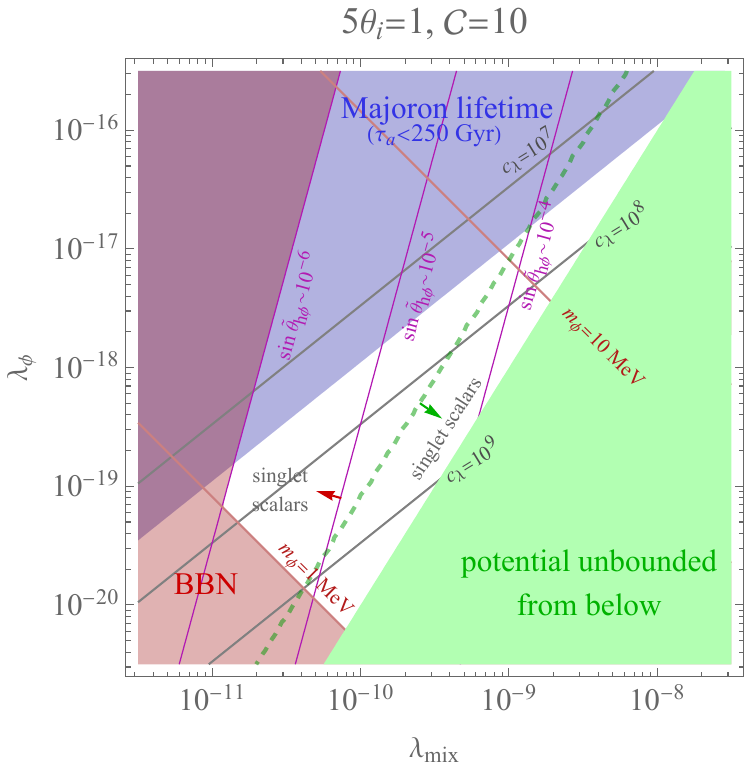}
    \caption{Constraints (shaded region) and key quantities (solid lines) are depicted in the plane of $\lambda_{\rm mix} = \lambda_{h \phi} + \lambda_{h s}/4$ and $\lambda_\phi$ for the Majoron model.
    The green-shaded region is excluded independent of the pNGB interactions due to potential instability from negative mixed quartics.
    The blue-shaded region is excluded by the lower bound of the Majoron lifetime (CMB and BAO)\,\cite{Audren:2014bca,Enqvist:2019tsa,Nygaard:2020sow,Alvi:2022aam,Simon:2022ftd}, taking 
    $\sum m_\nu^2 \lesssim (0.05\,\eV)^2$.
    The red region shows BBN constraints\,\cite{Fradette:2017sdd}, where $\phi$ decays too late ($m_\phi<2m_e$  or $\sin \theta_{h\phi}$  too small).
    Solid lines indicate benchmarks: $m_\phi$ (red), $c_\lambda$ (gray), 
    $\sin \tilde \theta_{h\phi}$ (purple).
    Arrowed regions, excluded in the minimal Higgs portal model, become viable with additional singlet scalars.
    }
    \label{Fig:final}
\end{figure}

Lepton asymmetry is generated via lepton-number-violating processes, where the dominant contributions are from the decay and inverse decay of RHNs\,\cite{Chun:2023eqc}.
Note that they used to be wash-out terms in thermal leptogenesis, but now they are the ones generating lepton asymmetry up to the equilibrated value in the presence of $\dot \theta \neq 0$.
The lepton number then gets transported into the baryon number via electroweak sphaleron, and the equilibrated value of the baryon number density is given by $n_B = c_B \,\dot \theta \,T^2/6$ with $c_B\simeq \frac{7}{33}$.

Thus, we need both electroweak sphaleron and RHNs in the thermal bath at high temperatures.
Since the masses of RHNs are given by $M_{N_i}(T) = y_i f_a(T)/\sqrt{2} \simeq y_i \sqrt{c_\lambda} T/\sqrt{2}$ at $T>T_c$, we need $y_i \lesssim 1/\sqrt{c_\lambda} = \sqrt{3\lambda_\phi/\lambda_{\rm mix}}$.
For simplicity, we take a benchmark choice, $y_i = \sqrt{\lambda_{\rm mix}}$, which leads to $M_{N_i}(T)/T = \sqrt{\lambda_{\rm mix}^2/6\lambda_\phi}$ less than one because of the stability condition $\lambda_{\rm mix}^2<\lambda_\phi \lambda_s<\lambda_\phi$ of the $\phi$ potential. 
The following discussion is insensitive to $O(1)$ changes of $y_i$ and $Y_{D_{\alpha i}}$ unless RHNs are nearly degenerate or some of them are long-lived.

The baryon asymmetry is effectively frozen when either the electroweak sphaleron is decoupled at $T_{\rm EW} \simeq 130\,\GeV$\,\cite{DOnofrio:2014rug, Kuzmin:1985mm} or RHNs disappear via their decay.
The latter case could happen around $T=M_N^{(0)}/z_{\rm fo}$ after $M_N(T)$ gets saturated to $M_N^{(0)}\simeq y f_a^{(0)}/\sqrt{2}$. 
As will be shown later, our allowed parameter space implies that RHNs have to be lighter than a few hundred $\GeV$, so the electroweak sphaleron gets decoupled first.
Then, baryon asymmetry is frozen at $T_{\rm EW}$ and we obtain
\bal
Y_B
= \!
\frac{45 c_B}{2\pi^2 g_*} \bigg(\frac{\dot \theta}{T} \bigg)_{\rm \!\! slide} \!\!\!\!\!\!\! \times
\begin{cases}
    1 & \text{for $T_{\rm EW}>T_c$}
    \\
    \big(\frac{T_{\rm EW}}{T_c} \big)^2
     & 
     \text{for $T_{\rm EW}<T_c$}
\end{cases}
\label{Eq:YB}
\eal
where the suppression factor in the $T_{\rm EW}<T_c$ case comes from $T^3$ scaling of $\dot \theta$ during $T_{\rm osc}<T<T_c$. 
We find that $T_{\rm EW}<T_c$ is indeed the case in our allowed parameter space, so we assume $T_{\rm EW}<T_c$ in the following discussion for simplicity.
This means that baryon asymmetry is frozen at $T_{\rm EW}$.
We also check that it is sufficiently higher than the fragmentation temperature, and the required Dirac Yukawa coupling becomes $Y_D  \sim 10^{-7}$, considering $y_i^2\sim \lambda_{\rm mix}$.

Using Eqs.\,\eqref{Eq:thdotmax}, \eqref{Eq:DM} and \eqref{Eq:YB}, we finally obtain
\bal
&m_a^{(0)}= \frac{5\,\eV }
{{\cal C}^{\frac{1}{9}} (5\theta_i)^{\frac{4}{9}}}
\bigg( \! \frac{g_*}{100} \! \bigg)^{\!\! \frac{1}{3}} 
\bigg( \! \frac{10^8}{c_\lambda} \! \bigg)^{\!\! \frac{5}{9}} ,
\label{Eq:ma0}
\\
&f_a^{(0)} = {3\times 10^6\,\GeV }\,
\mathcal{C}^{\frac{1}{18}}
{(5\theta_i)^{\frac{2}{9}}}
\bigg( \! \frac{100}{g_*} \! \bigg)^{\!\! \frac{1}{6}} \!
\bigg( \! \frac{c_\lambda}{10^8} \! \bigg)^{\!\! \frac{5}{18}}
.
\label{Eq:fa0}
\eal
Assuming these estimations of $m_a^{(0)}$ and $f_a^{(0)}$, Fig.\,\ref{Fig:final} shows the viable parameter space in the $\lambda_{\rm mix}$-$\lambda_\phi$ plane with $c_\lambda = \lambda_{\rm mix}/3\lambda_\phi$, fixing $5\theta_i = \mathcal{O}(1)$ and $M_N^{(0)}= \sqrt{\lambda_{\rm mix}} \,f_a^{(0)}/\sqrt{2}$.

Shaded regions show the relevant constraints.
The region shaded by green corresponds to $\lambda_{\rm mix}^2 > \lambda_\phi \lambda_X$ with $X=h$ or $s$ where the potential is unbounded from below due to the large negative mixed quartic with $\lambda_X$ being the self coupling of $X$ that generates the negative mass-squared correction.
We take $\lambda_X=1$ to estimate this constraint while in the minimal model, $X$ is the SM Higgs whose self-quartic coupling is given by $\lambda_h\simeq 1/8$, so the constraint becomes as strong as the green dashed line.

The blue-shaded region is excluded by the lifetime bound of the Majoron dark matter.
It comes from analyses of the cosmic microwave background (CMB) and the baryon acoustic oscillation (BAO)\,\cite{Audren:2014bca,Enqvist:2019tsa,Nygaard:2020sow,Alvi:2022aam,Simon:2022ftd}, whereas we take the current upper bound of neutrino masses $\sum m_\nu^2 \simeq (0.05\,\eV)^2$.
This lifetime constraint can become stronger in $\lambda_\phi$-$\lambda_{\rm mix}$ plane by future neutrino mass measurements.

BBN also puts a strong constraint on $\phi$ as depicted by the red-shaded region\,\cite{Fradette:2017sdd}.
In this region, $\phi$ decays after BBN either because $m_\phi$ is smaller than the electron threshold $2m_e\simeq \MeV$, or $\sin \theta_{h\phi}$ is too small.
Note that $\sin \theta_{h\phi}$ cannot be fully determined in this plane because $\lambda_{h\phi} = \lambda_{\rm mix} - N_s \lambda_{\phi s_i}/4$ depends on $\lambda_{\phi s_i}$ for a given $\lambda_{\rm mix}$.
Assuming $\lambda_{h\phi}\sim N_s \lambda_{\phi s_i}/4 \sim \lambda_{\rm mix}$, we naively estimate the mixing angle as $\tilde \theta_{h\phi} \sim 2\lambda_{\rm mix} f_a^{(0)} v_h/m_h^2$, and purple lines depict some of its benchmark values, $10^{-4}$, $10^{-5}$, and $10^{-6}$.
For the minimal Higgs portal model, we expect $\tilde \theta_{h\phi} =  \theta_{h\phi}$, and therefore, the region left to $\sin \tilde \theta_{h\phi} = 10^{-5}$ is excluded, but it can be rescued by additional scalars with $10\,\%$-level tuning between $\lambda_{h\phi}$ and $\lambda_{\phi s_i}$.
For $\sin \tilde \theta_{h\phi} \gsim 10^{-4}$, we have chances to prove our model via kaon experiments.

Other solid lines also represent values of relevant physical quantities.
Red lines correspond to $m_\phi$, showing that $m_\phi \lsim 20\,\MeV$ in our parameter space, and gray lines represent several values of $c_\lambda$. 

The allowed parameter space indicates that $m_\phi(T) \sim \sqrt{\lambda_\phi c_\lambda} T$ for $T>T_c$ is greater than the Hubble rate for $T\lesssim \sqrt{\lambda_\phi c_\lambda} M_{\rm Pl}$.
Therefore, the time scale of $\phi$ dynamics is short enough, and its oscillation effects on the $\theta$ dynamics should be averaged out, which justifies $\phi=f_a$ treatment in Eq.\,\eqref{Eq:eom}.
Moreover, the $\phi$ oscillation gets quickly damped out by the thermal friction of $\Gamma \sim y^2 T$ with $y\sim 10^{-4}$, which makes $\phi$ adiabatically follow $f_a$.

Although it is not depicted in the figure, the Eqs.\,\eqref{Eq:ma0} and \eqref{Eq:fa0} have tension with the isocurvature perturbation constraint from CMB observation\,\cite{Planck:2018jri} with taking reheating temperature greater than $T_0$ in Eq.\,\eqref{eq:T0}.
However, this can be easily avoided by introducing a small negative Hubble-induced mass\,\cite{Stewart:1994ts, Dine:1995kz, Dine:1995uk} or coupling to the inflaton\,\cite{Chun:2014xva}, making $f_a$ large during inflation without changing our scenario after reheating.
As it is an independent sector, we leave detailed studies for future work.

In conclusion, this is an explicit example model for $n=5$ with testable predictions: (i) $\phi$ with $\MeV < m_\phi \lsim 20\,\MeV$ and $10^{-6}\lsim \sin \theta_{h\phi} \lsim 10^{-4}$, (ii) a RHN with $0.1\,\GeV \lsim M_N^{(0)} \lsim 100\,\GeV$, and (iii) Majoron DM with $m_a^{(0)} \sim \eV$ and $f_a^{(0)} \sim 10^6\,\GeV$.
In future, it is expected to test Higgs portal mixing up to $\sin \theta_{h \phi}\lsim 4\times 10^{-5}$ via KOTO step-2\,\cite{Goudzovski:2022vbt, Nomura:2020oor, Aoki:2021cqa} and KLEVER\,\cite{Goudzovski:2022vbt, Beacham:2019nyx}, weak-scale RHNs at future colliders \,\cite{Chun:2019nwi, Abdullahi:2022jlv}, and weak-scale singlet scalars at LHC\,\cite{CMS:2024zqs, ATL-PHYS-PUB-2022-007, ATL-PHYS-PUB-2023-008, ATL-PHYS-PUB-2023-018} and future colliders\,\cite{Bernardi:2022hny, Narain:2022qud, ILCInternationalDevelopmentTeam:2022izu, ILC:2013jhg, Bambade:2019fyw, Linssen:2012hp, CLICdp:2018cto, FCC:2018evy, CEPCStudyGroup:2018ghi, CEPCPhysicsStudyGroup:2022uwl, FCC:2018vvp}. 
However, the eV-scale Majoron DM evades constraints applicable for thermally produced Majorons\,\cite{Sabti:2019mhn,Blinov:2019gcj,Sandner:2023ptm,Chang:2024mvg}, since thermal abundance remains suppressed for $c_{\lambda} \gtrsim 10^{7}$. The growth of fluctuations can have observational consequences in small-scale structures\,\cite{Eroncel:2022efc}.

\section{Discussion} 
Now, let us discuss the model-building aspect of the explicit breaking operator.
In the following, we show how our dimension-five ($n=5$) explicit breaking setup can be realized in a technically natural setup. 
Then, we discuss the generalization of our mechanism to an $n\geq 6$ setup.

In our $n=5$ example, our working parameter space indicates a theoretically peculiar condition; the cutoff scale $\Lambda \equiv \frac{25}{2\sqrt{2}} \frac{f_a^3}{m_a^2} $ has to be much larger than the Planck scale.
This condition looks peculiar because (i) it seems contradictory to the fact that quantum gravity does not respect any global symmetry, and (ii) higher-dimensional operators with smaller cutoff can be more important than this five-dimensional operator.
However, this setup can easily be realized by a spontaneously broken discrete symmetry.

For example, we can consider a discrete symmetry $\mathbb{Z}_{5+4x}$ under which $\Phi$ is charged by $1$, and introduce an additional complex scalar field $X$ whose charge is $x$.
Let us assume that $\mathbb{Z}_{5+4x}$ is exact since quantum gravity does not necessarily break discrete symmetries.
Then, the least-dimensional explicit breaking operator of the $U(1)$ becomes $X^4 \Phi^5/M_{\rm Pl}^{5}$.
If $\langle X \rangle \neq 0$, this operator induces the $\Phi^5/\Lambda$ operator, where $\Lambda = M_{\rm Pl}^5/\langle X \rangle^4$. 
Thus, a large $\Lambda$ can be realized if $\langle X \rangle \ll M_{\rm Pl}$; numerically, we need $\langle X \rangle \sim 10^{13}\,\GeV$.
The discrete symmetry enforces the next least-dimensional operator to be $\Phi^{5+4x}/M_{\rm Pl}^{1+4x}$ whose effect is subleading in the field range of $|\Phi| < \left( \frac{M_{\rm Pl}}{\Lambda} \right)^{\!\!1/4x} \!\!M_{\rm Pl}$, which is satisfied in our scenario if $x \geq 3$.
We also checked that the temperature-dependent vev of $\Phi$ does not ruin $\langle X \rangle$.
Therefore, our setup of the dimension-five explicit breaking operator can be realized in a technically natural setup (where $U(1)$ is accidentally protected by $\mathbb{Z}_{5+4x}$), which cannot be forbidden either theoretically or phenomenologically.

$n=5$ is actually not a necessary ingredient for our mechanism, although its cosmological scenario is simpler than the $n \geq 6$ case.
For $n \geq 6$, $m_a(T) \propto T^{\frac{n-2}{2}}$ decreases with temperature no slower than $H(T)\propto T^2$, which may imply the pNGB's saturation at the potential minimum at high temperatures, in contrast to $n=5$.
However, this is not true. 
This case requires a more careful consideration of how the effective potential of $\phi$ is developed in cosmological history.

The large global minimum $f_a(T)\simeq \sqrt{c_\lambda}\,T$ of $\phi$'s effective potential requires the SM Higgs $h$ or singlets $s_i$ to be in the background plasma. 
However, they are not thermalized before the end of inflation. 
Depending on the production rate, they may not be thermalized even during/after reheating until the expansion rate becomes comparable to the production rate.
Therefore, it is natural to start with a zero-temperature potential and assume the pNGB remains frozen initially. 
Then, as the scalar particles (that have the negative mixed quartic with $\Phi$) are populated, the minimum of the effective potential gets shifted toward $f_a(T)$.
After these particles are fully thermalized, the potential minimum evolves as $f_a(T)$.

During the effective potential's deformation, $\phi$ rolls down toward the new minimum.
The pNGB mass $m_a$ increases as $\phi^{(n-2)/2}$, 
so at some point, it becomes greater than the Hubble rate (for $n\geq 6$) and starts its dynamics\,\footnote{
For $n=5$, at high temperatures, the pNGB mass is still smaller than the Hubble rate even after the complete thermalization.
This was the reason why we could separate the thermalization process from the evolution of pNGB.
}. 
As $f_a(T)\propto T$ decreases, the barrier height of the pNGB now decreases even faster compared to the case with $n=5$. 
Hence, the kinetic energy becomes higher than the potential barrier, enabling similar sliding dynamics. 
Note that the $\frac{\dot{\theta}}{T}={\rm constant}$ phase still exists, as long as $\frac{\dot{f_a}}{f_a}=-H$, leading to $\ddot{\theta}+H\dot{\theta}\simeq 0$, and the rest part of the scenario remains unchanged.
At $T<T_c$, the mass saturates to a constant value, and the pNGB gets trapped at a lower temperature and may play the role of DM.

Thus, our mechanism of pNGB sliding with symmetry non-restoration can extend beyond $n=5$ to higher-dimensional operators with additional parameters describing thermalization details at high temperatures. 
We leave a careful investigation and model building of this intriguing possibility to future work.

\section{Summary}
In this work, we have proposed a new mechanism of generating sliding pNGB dynamics by using the symmetry non-restoration setup.
Focusing on the case that the pNGB mass originates from an explicitly broken global U(1) symmetry by a five-dimensional operator, we have shown that an initial misalignment of a pNGB can successfully generate the observed baryon asymmetry as well as dark matter abundance. 

The symmetry non-restoration is realized by additional scalar fields that have negative mixed quartic coupling to the global symmetry-breaking field, generating a negative thermal mass-squared correction, which in turn makes the symmetry-breaking scale larger at high temperatures.
Consequently, the mass of the pNGB is also enhanced at high temperatures.
As the temperature drops, the height of the pNGB potential, which is determined by the pNGB mass, also decreases. 
When the Hubble friction becomes comparable to the potential barrier, the pNGB starts its dynamics while the height of the barrier decreases faster than the redshift of the kinetic energy of the pNGB.
So, the pNGB begins sliding along the potential.

The pNGB sliding serves as an external chemical potential for lepton number, enabling the spontaneous leptogenesis mechanism.
The generated lepton asymmetry is converted to the baryon asymmetry via electroweak sphaleron, and the baryon asymmetry is frozen after the electroweak sphaleron is decoupled. 
After the potential height gets saturated below $T_c$, the kinetic energy starts to redshift in the exactly same way as the usual kination mode. 
Before getting eventually trapped at the `would-be' oscillation temperature, the fluctuations of the pNGB get significantly enhanced due to parametric resonance, leading to a complete fragmentation of the homogeneous mode. 
The final DM abundance thus exists in the form of pNGB fluctuations.

We have shown that this scenario can be realized in the extended type-I seesaw model where the right-handed neutrino masses are generated by the spontaneous breaking of the global $U(1)_{B-L}$.
Phenomenologically allowed parameter range is summarized in Fig.\,\ref{Fig:final} and Eqs.\,\eqref{Eq:ma0} and \eqref{Eq:fa0}.

Although we have focused on the dimension-five explicit breaking operator, we also show that our mechanism can be extended beyond $n=5$ to higher-dimensional operators, which requires a careful study of the full thermalization history. 
A detailed exploration of such scenarios is left for future work.

\vspace{0.2cm}
\noindent{\bf Acknowledgement}
We thank Takeo Moroi and Kyohei Mukaida for their useful comments.
This work was supported by IBS under the project code, IBS-R018-D1.

\appendix

\section{Realizing a large $c_\lambda$}\label{ApA}
\subsection{Model}
For simplicity, let us consider a scalar model
\bal
V(\Phi, S) \! &= \! \! \lambda_\phi |\Phi|^4 \! -\! 2 \lambda_{\phi s}|\Phi|^2 |S|^2 \! +\! \lambda_s |S|^4
\!-\! m_0^2 |\Phi|^2 \! +\! m_s^2 |S|^2 \!,
\eal
where $S$ is an additional complex scalar
with $\lambda_\phi,\, \lambda_{\phi s}, \, \lambda_s, \, m_0^2,\, m_s^2 >0$.
One can trivially replace $S$ by the SM Higgs doublet $H$ with proper treatments of Higgs vacuum expectation value (vev) and mass conditions and degrees of freedom.

Let us denote the radial modes of $\Phi$ and $S$ by $\phi$ and $s$, respectively.
Their potential is given by
\bal
V = \frac{1}{4} \lambda_\phi \phi^4 -\frac{1}{2} (m_0^2 + \lambda_{\phi s} s^2) \phi^2
+\frac{1}{4}\lambda_s s^4 + \frac{1}{2} m_s^2 s^2.
\eal
At each slice of constant $s$, $V$ is minimized at $\phi^2 = (m_0^2 + \lambda_{\phi s} s^2)/\lambda_\phi$.
Putting this back to the potential gives the potential along this trajectory,
\bal
V \to  -\frac{(m_0^2 + \lambda_{\phi s} s^2)^2}{4\lambda_\phi} + \frac{1}{4}\lambda_s s^4 + \frac{1}{2} m_s^2 s^2.
\eal
For the stability of the potential at large $s$, we need the effective quartic term to be positive, which can be written as
\bal
\lambda_\phi \lambda_s - \lambda_{\phi s}^2 >0.
\label{Eq_S:stability}
\eal
In addition, to make $\langle s \rangle =0$ (although it is not necessary), we require the quadratic term to be positive, so
\bal
m_s^2 > \frac{\lambda_{\phi s} }{\lambda_\phi} m_0^2
\sim \lambda_{\phi s} \, {f_a^{(0)}}^2.
\label{Eq_S:ms_condition}
\eal
For the Higgs case, one needs to solve the conditions of $v_h=246\,\GeV$ and $m_h=125\,\GeV$, which fixes the quadratic and quartic couplings as functions of $f_a^{(0)}$ and $\lambda_{h\phi}$.
However, these conditions may not be sufficient, since the potential can be spoiled by radiative corrections.
In the next subsection, we will check self-consistency numerically by including the Coleman-Weinberg potential, which is given by
\bal
&V_{\rm CW} \!\! = \!\! \frac{1}{64\pi^2} \! \Bigg[ \!
(3\lambda_\phi \phi^2 \!\! -\!\! \lambda_{\phi s} s^2 \!\! -\! m_0^2)^{\! 2} \! \Big( \! \log \! \frac{\! 3\lambda_\phi \phi^2 \!\!-\!\!\lambda_{\phi s} s^2 \!\!-\! m_0^2}{\mu^2} \!-\! \frac{3}{2} \Big)
\nn\\
&
+
(\lambda_\phi \phi^2 -\lambda_{\phi s} s^2-m_0^2)^2 \Big( \log \frac{\lambda_\phi \phi^2 -\lambda_{\phi s} s^2-m_0^2}{\mu^2}-\frac{3}{2} \Big)
\nn
\\
&
+
(3\lambda_s s^2 \!\! -\lambda_{\phi s} \phi^2 +m_s^2)^2 \Big( \log \frac{3\lambda_s s^2-\lambda_{\phi s} \phi^2 +m_s^2}{\mu^2}-\frac{3}{2} \Big)
\nn\\
&
+
(\lambda_s s^2-\lambda_{\phi s} \phi^2 +m_s^2)^2 \Big( \log \frac{\lambda_s s^2-\lambda_{\phi s} \phi^2 +m_s^2}{\mu^2}-\frac{3}{2} \Big)
\Bigg].
\label{Eq_S:VCW}
\eal
In order to keep our analysis determined by the tree-level potential, we need $V_{\rm CW}$ to be small compared to the tree-level.
For this, we need $\lambda \gg \lambda' \lambda''/16\pi^2$ (with $\lambda, \lambda', \lambda'' = \lambda_\phi, \lambda_{\phi s}$ or $\lambda_s$).
Once it is satisfied, one-loop contributions become sufficiently small compared to the tree-level potential except for the $\lambda_{\phi s} m_s^2 \phi^2$ term from the third and fourth lines of \eqref{Eq_S:VCW}.
Therefore, the vev of $\phi$ should be estimated with
\bal
m_\phi^2 &=
m_0^2 + \frac{\lambda_{\phi s}}{16\pi^2} m_s^2 \log \frac{m_s^2}{\mu^2}
\\
\langle \phi \rangle &\equiv f_a^{(0)}
\simeq \frac{m_\phi}{\sqrt{\lambda_\phi}},
\eal
which brings $m_s$ to be as small as possible unless we finely tune the parameters $m_0^2$ and $m_s^2$ (this tuning may be acceptable since we anyway ignore all the UV corrections to scalar masses).
When we take the smallest value of \eqref{Eq_S:ms_condition}, the correction to $m_\phi^2$ is roughly $\delta m_\phi^2/m_\phi^2 \sim O(\lambda_{\phi s}^2/(16\pi^2 \lambda_\phi))\lsim O(1)$.

\begin{figure*}
    \centering
    \includegraphics[width=0.48\textwidth]{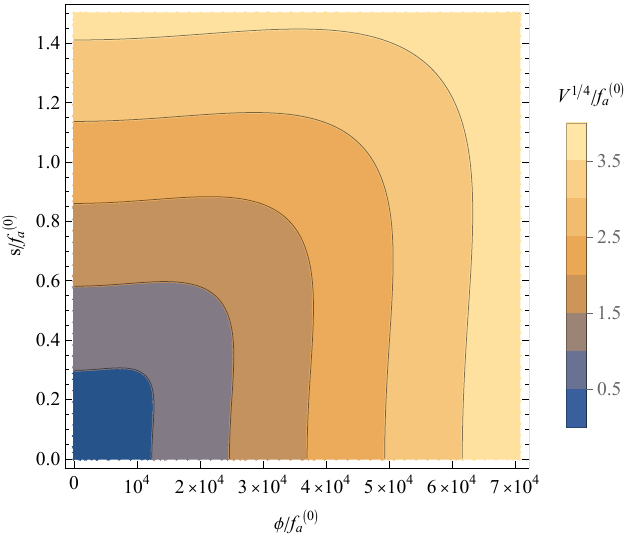}
    \includegraphics[width=0.42\textwidth]{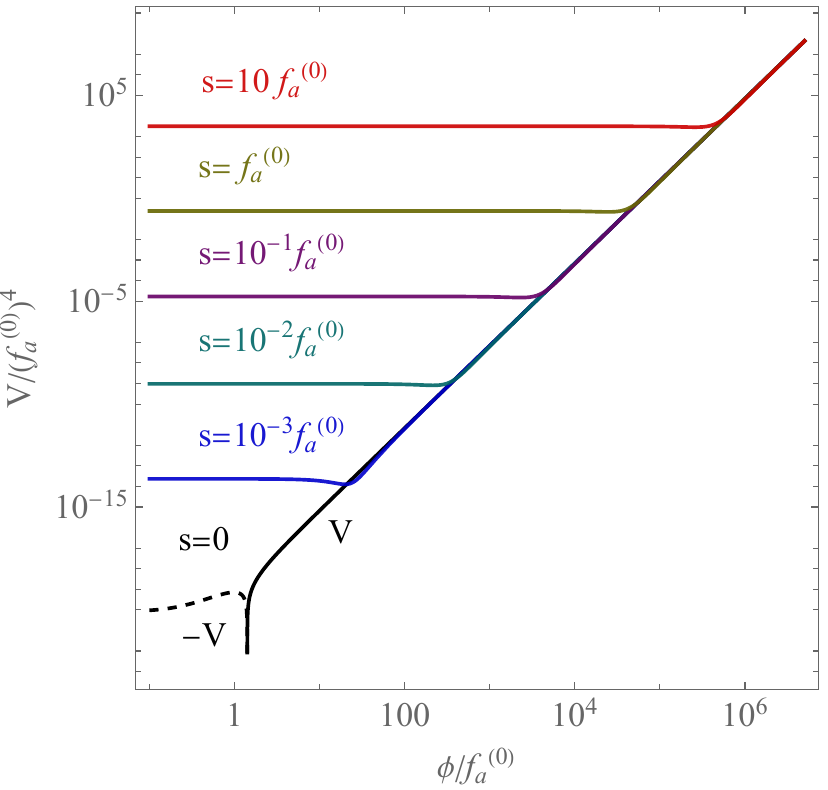}
    \caption{[Left] One-quarter power of the zero temperature potential is depicted in $\phi$-$s$ space. As it is shown, there is no unstable direction. [Right] The potential in the $\phi$ direction is depicted for fixed $s/f_a^{(0)} = 10^{-3}, \, 10^{-2}, \, 10^{-1},\, 1,$ and $10$. }
    \label{Fig:stability}
\end{figure*}

With $\langle s \rangle =0$, we have the temperature-dependent potential
\bal
V_T(\phi) &= \frac{\lambda_\phi}{4}\phi^4 - \frac{m_0^2}{2} \phi^2 + \left. V_{\rm CW}\right|_{s=0}
\nn
\\
&
+ \frac{T^4}{2\pi^2}J_B \left( \frac{3\lambda_\phi \phi^2 -m_0^2}{T^2} \right)
+ \frac{T^4}{2\pi^2}J_B \left( \frac{\lambda_\phi \phi^2 -m_0^2}{T^2} \right)
\nn
\\
&+ 2\cdot \frac{T^4}{2\pi^2}J_B \left( \frac{m_s^2-\lambda_{\phi s} \phi^2}{T^2} \right),
\eal
where $J_B$ is the thermal correction from a bosonic degree, defined as
\bal
    J_{B}(m^2/T^2) &= \int_0^{\infty} dx\, x^2 \log\left( 1 - e^{-\sqrt{x^2+m^2/T^2}} \right),
\label{Eq_S:VT}
\eal
and its high-$T$ expansion can be written as
\bal
J_B(m^2/T^2)
    &\simeq
    -\frac{\pi^4}{45}
    +\frac{\pi^2}{12} \frac{m^2}{T^2}
    +\cdots.
\eal
Therefore, at a high temperature with $T\gg m_s$ and $T \gg \sqrt{\lambda_{\phi s}} \phi$, we can expand Eq.\,\eqref{Eq_S:VT} as
\bal
&V_T(\phi) \simeq \frac{\lambda_\phi}{4}\phi^4 - \frac{m^2}{2} \phi^2 +V_{\rm CW}
\\
&
    - \!4 \! \cdot \! \frac{\pi^2}{90}T^4
    \!\! +\! \frac{T^2}{24} \!\bigg(
        \! (3\lambda_\phi \phi^2 \!\! - \! m^2 )
        \! +\! (\lambda_\phi \phi^2 \!\! - \! m^2 )
        \! + \! 2(m_s^2 \!\!- \! \lambda_{\phi s} \phi^2 )
        \! \bigg)
\nn
\\
&\simeq
\frac{\lambda_\phi}{4}\phi^4
- \frac{1}{2}
\bigg(m_\phi^2+\frac{\lambda_{\phi s}-2\lambda_\phi }{6}T^2 \bigg) \phi^2 + \cdots.
\eal
In the final expression, we have ignored constant terms and the Coleman-Weinberg potential except for the quadratic correction.
Therefore, we obtain
\bal
c_\lambda = \frac{N_s\lambda_{\phi s}-4\lambda_\phi }{12\lambda_\phi},
\label{Eq_S:clambda}
\eal
where $N_s$ is the degree of freedom of $S$.
When $S$ is a complex scalar field as we set here, $N_s=2$ while $N_s=4$ should be taken if one replaces $S$ by the SM Higgs doublet $H$.
Note that, to have $c_\lambda>0$, we need $\lambda_{\phi s} > 2\lambda_\phi$.

On the other hand, $S$ has to be in the thermal bath to provide thermal corrections.
If $S$ were the Higgs boson, we do not need to introduce additional interactions, but for $S$, this can be realized by a mixed quartic coupling between $S$ and the SM Higgs doublet $H$,
\bal
\lambda_{hs} |H|^2 |S|^2,
\label{Eq_S:lambda_hs}
\eal
from which the production rate of $S$ is given by $\Gamma \sim \lambda_{hs}^2 T$.
Having it to be greater than the Hubble rate at $T\simeq T_{\rm EW}$ ($T_0$) provides a lower bound of $\lambda_{hs} \gsim 10^{-8}$ ($10^{-3}$).
Of course, one can introduce other interactions instead of \eqref{Eq_S:lambda_hs}.

\subsection{Numerical study}

To realize a large $c_\lambda$, we need $\lambda_{\phi s} \gg \lambda_\phi$, but we have to satisfy the stability condition\,\eqref{Eq_S:stability}, which can be rewritten as
\bal
c_\lambda \simeq
\frac{\lambda_{\phi s}}{6\lambda_\phi} < \frac{\lambda_s}{6\lambda_{\phi s}},
\eal
so we have to take a large $\lambda_s$ and a small $\lambda_{\phi s}$.
At the same time, $\lambda_\phi$ must be $O(\lambda_{\phi s}^2)$ to maximize $c_\lambda$.

Here, we chose benchmark parameters to provide $c_\lambda \simeq 10^8$;
\bal
&\lambda_{\phi s} = \frac{1}{6}\times 10^{-9},
\quad
\lambda_\phi = 10 \lambda_{\phi s}^2,
\quad
\lambda_s = 1,
\nn
\\
&m_s = \sqrt{2\lambda_{\phi s}} f_a^{(0)},
\quad
m_0 = \sqrt{\lambda_\phi} f_a^{(0)}
\eal
Note that changing these parameters to make $c_\lambda > 10^8$ is not difficult at all.
We take $\mu=f_a^{(0)}$ in the following analysis.

First, we check that $\langle s \rangle =0$, and the zero-temperature potential is not spoiled by quantum corrections.
We include the full expression of the Colemann-Weinberg correction given in Eq.\,\eqref{Eq_S:VCW} and depict it in Fig.\,\ref{Fig:stability}.
$V^{1/4}$ is depicted in $\phi$-$s$ space in the left panel, and the potential in the $\phi$ direction is depicted for fixed $s/f_a^{(0)} = 10^{-3}, \, 10^{-2}, \, 10^{-1},\, 1,$ and $10$.
As shown in the figures, the potential is minimized at $\langle \phi \rangle \simeq f_a^{(0)}$, and there is no minimum at nonzero $s$.

\begin{figure*}
    \includegraphics[width=0.48\textwidth]{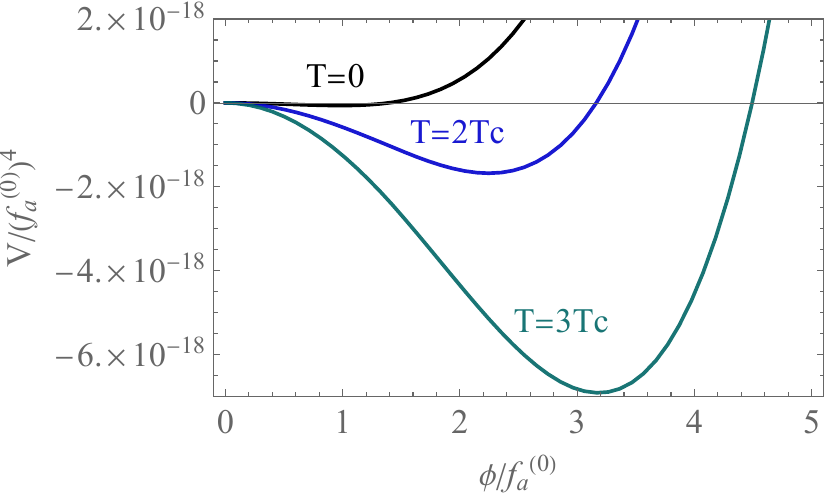}
    \includegraphics[width=0.42\textwidth]{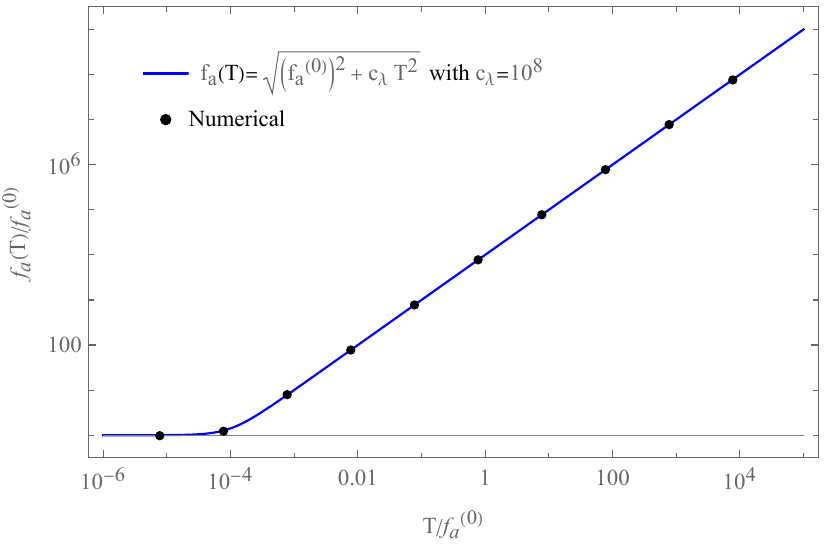}
    \caption{[Left] $V(\phi)$ with $s=0$ is depicted for different temperatures, $T=0, \, 2T_c,$ and $3T_c$.
    [Right] $f_a(T)=\langle \phi \rangle_T$ is depicted as a function of temperature. Black dots indicate the values we obtained by minimizing the full one-loop temperature-dependent effective potential while the blue line corresponds to the expected line from the approximated effective potential.}
    \label{Fig:VT}
\end{figure*}

With taking $s=0$, the temperature-dependent potential $V_T(\phi)$ is depicted in the left panel of Fig.\,\ref{Fig:VT}.
In the figure, we fix $V(0)=0$ by a constant shift to make a comparison among different temperatures.
We scan temperatures and find $\langle \phi \rangle_T$ at each temperature.
Our numerical finding is shown by black dots in the right panel of Fig.\,\ref{Fig:VT}, which make good agreement with our analytic expression $f_a(T) \simeq \sqrt{(f_a^{(0)})^2 + c_\lambda T^2}$ with taking $c_\lambda = 10^{8}$ as depicted by the solid blue line.

\section{Numerical study of \lowercase{p}NGB dynamics}\label{app-num}\label{ApB}

\begin{figure*}
    \centering
    \includegraphics[width=0.48\textwidth]{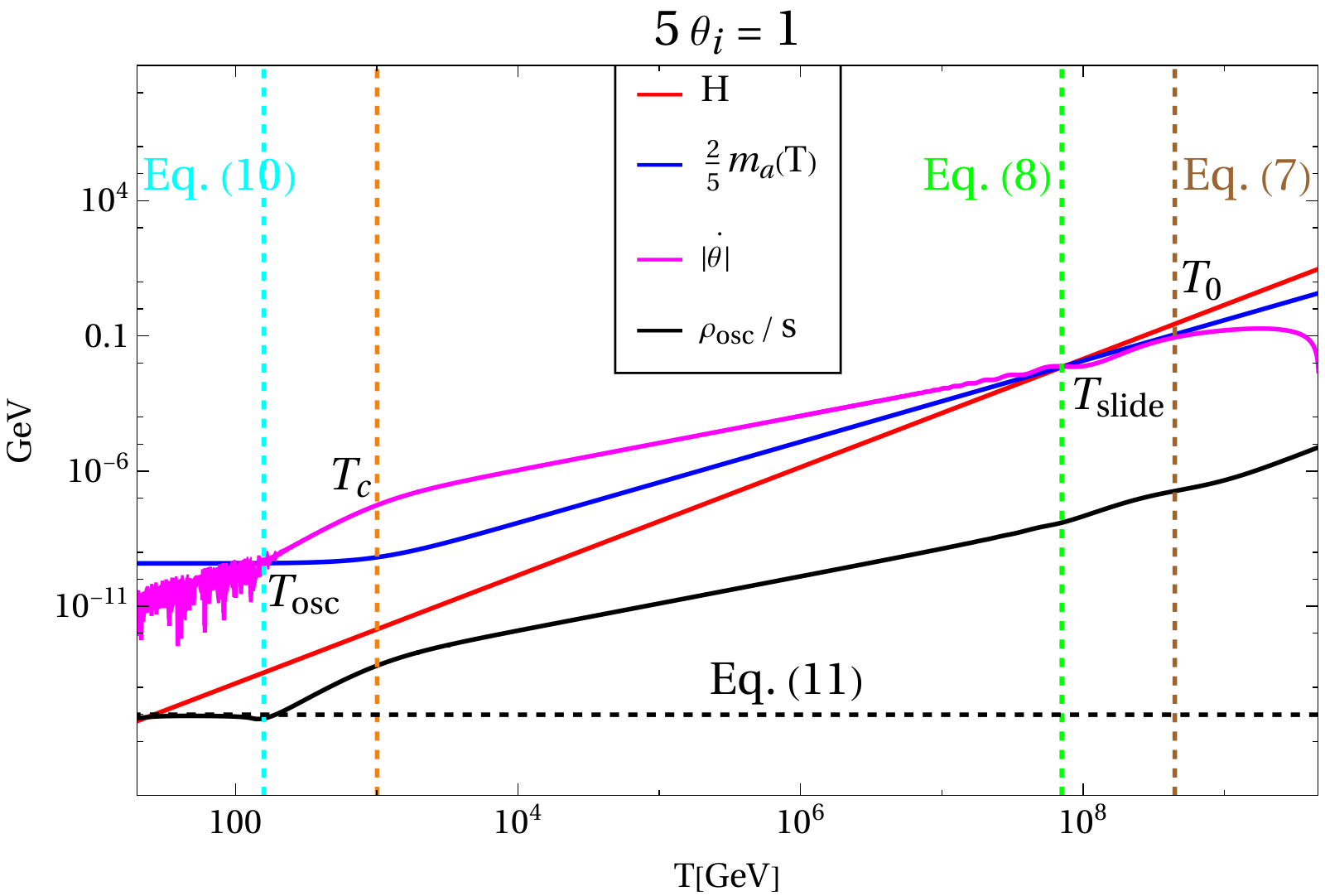}
 \includegraphics[width=0.48\textwidth]{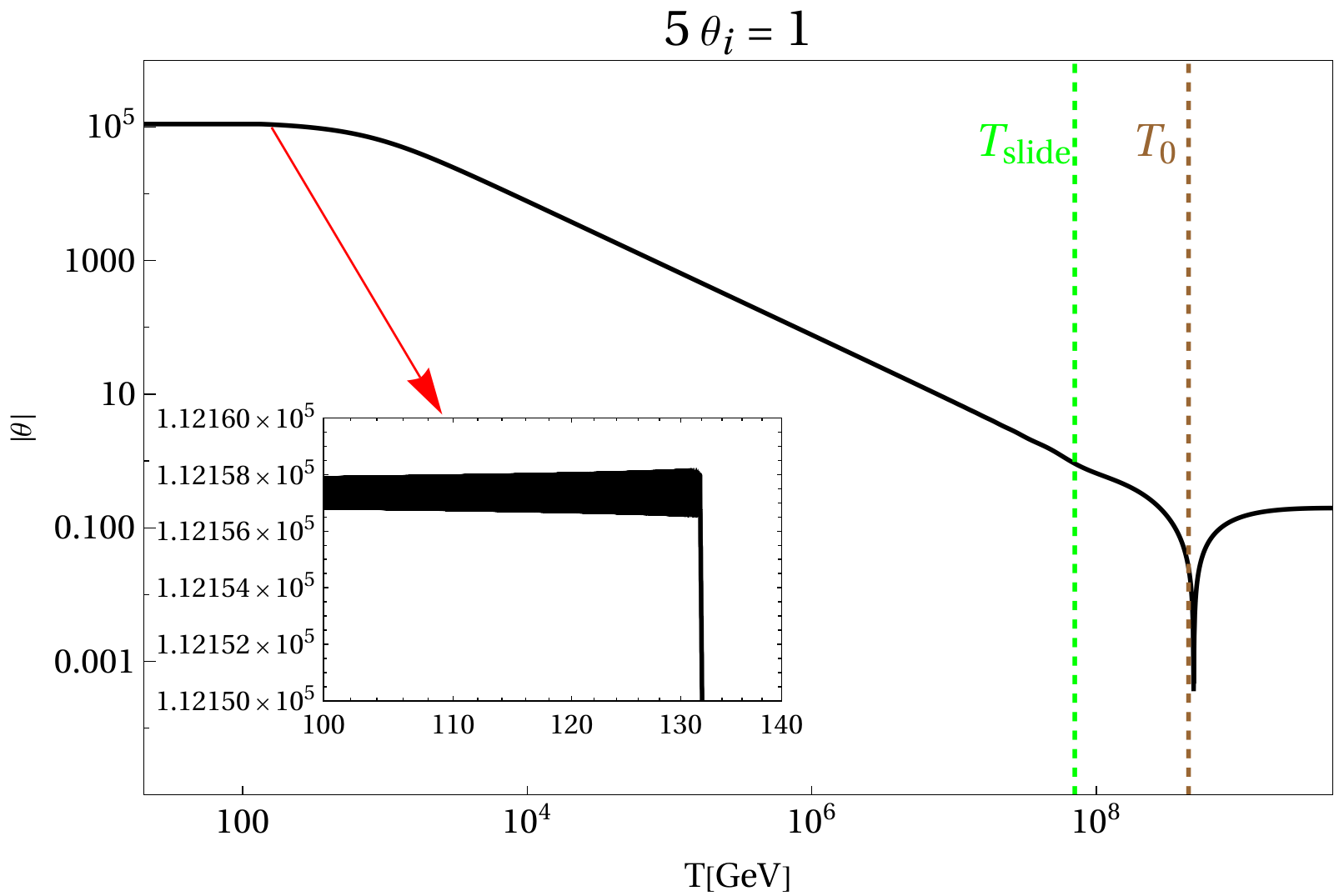}
    \caption{Left panel: Evolution of $H$, $\frac{2}{5}m_a (T)$, $|\dot{\theta}|$ and $\rho_{\rm osc} /s$ with temperature considering $f_a^{(0)} = 10^{6} ~\text{GeV}$, $m_a^{(0)} = 1 ~ \text{eV}$, $c_{\lambda} = 10^6$ and $5 \theta_i =1$. Right panel: Evolution of $|\theta|$ for the same parameters.}
    \label{Fig:Evolution}
\end{figure*}

\begin{figure*}
    \centering
    \includegraphics[width=0.48\textwidth]{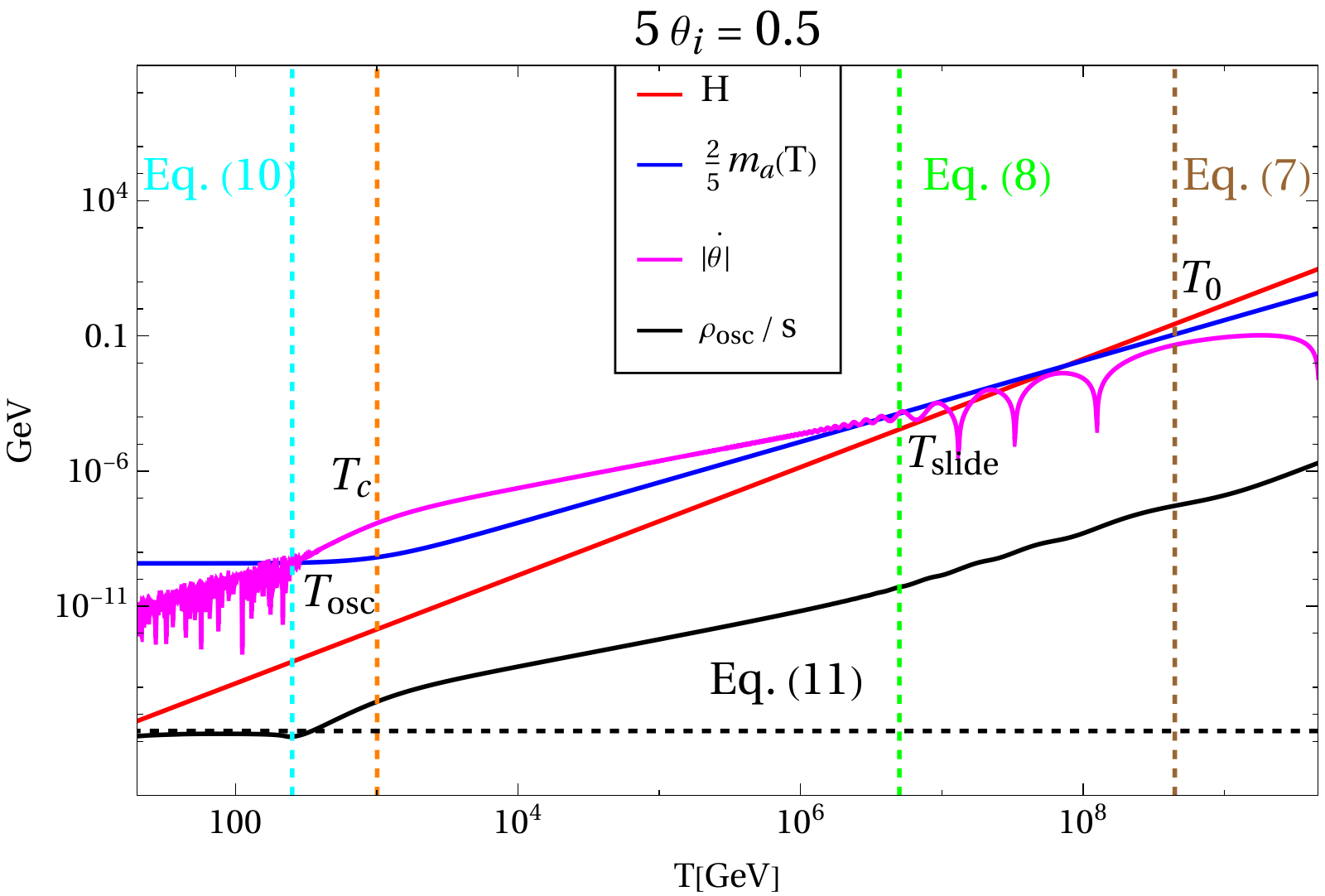}
    \includegraphics[width=0.48\textwidth]{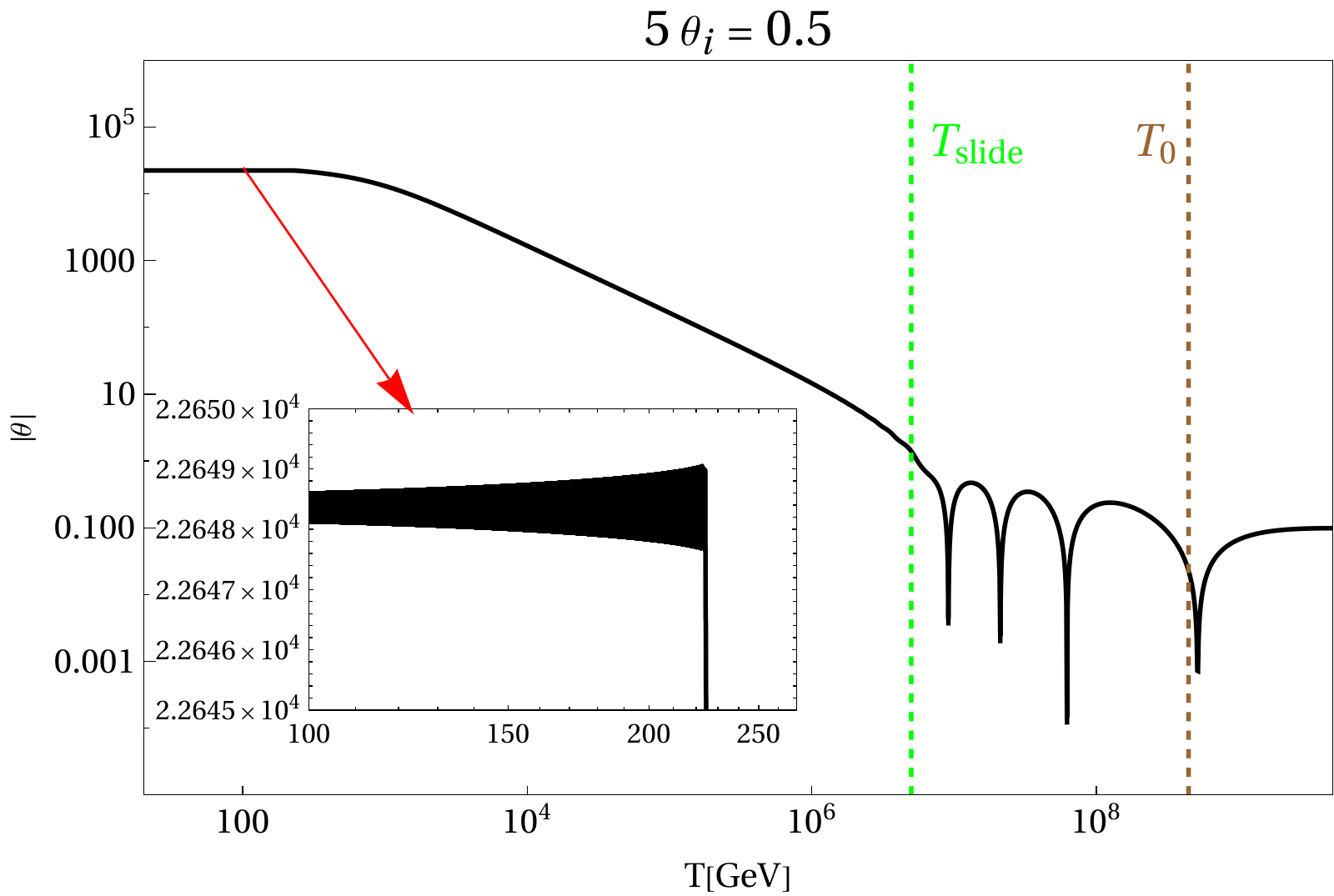}
    \caption{Same as in Fig. \ref{Fig:Evolution}, but for $5 \theta_i = 0.5$.}
    \label{Fig:Evolution2}
\end{figure*}

\begin{figure*}
    \centering
    \includegraphics[width=0.48\textwidth]{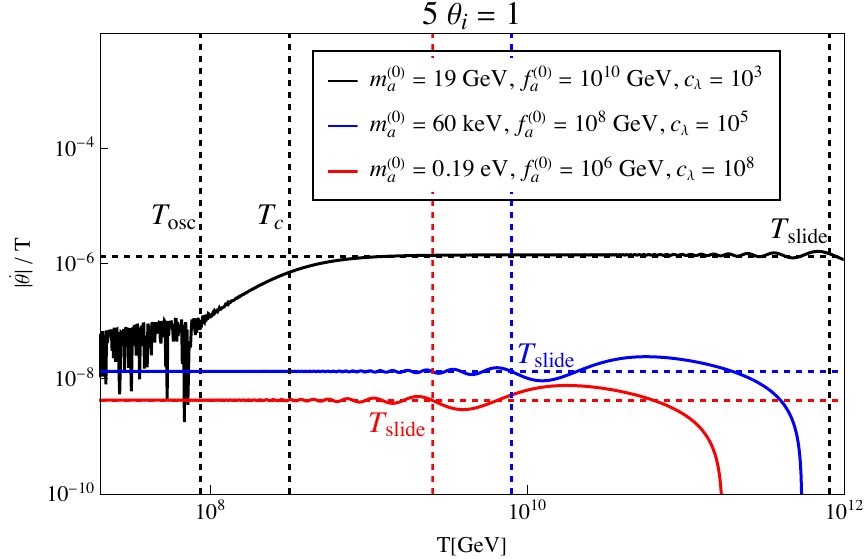}
    \includegraphics[width=0.48\textwidth]{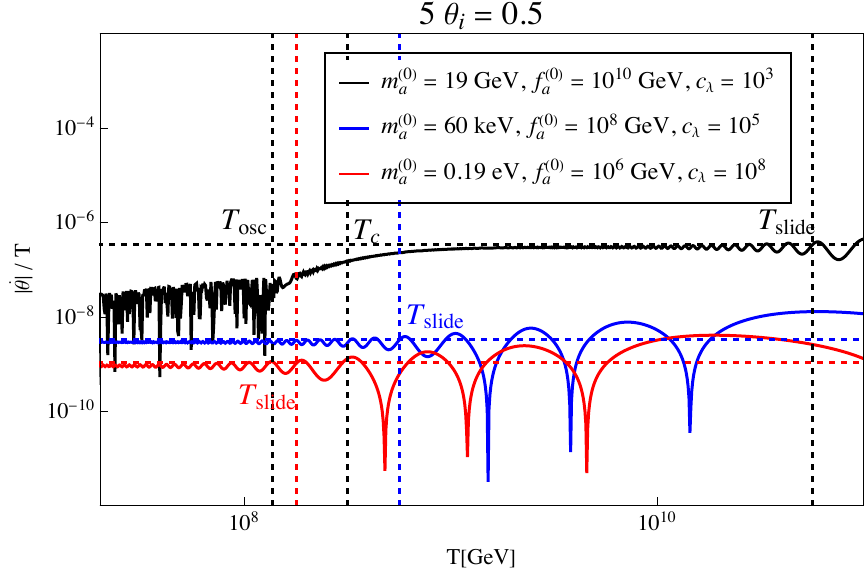}
    \caption{Evolution of $|\dot{\theta}|/ T$ for different benchmark parameters considering $5 \theta_i =1$ (left panel) and $5 \theta_i =0.5$ (right panel). The solid lines indicate the results from our numerical solutions while the dashed lines represent our analytical estimates.}
    \label{Fig:thetadot}
\end{figure*}

Here, we demonstrate the dynamics of $\theta$ in the different regimes by numerically solving its equation of motion (cf. Eq. \eqref{Eq:eom}) and check the consistency of our analytical estimates derived in the main text. We present our numerical results in Figs. \ref{Fig:Evolution}-\ref{Fig:thetadot}, choosing some benchmark parameters. In Fig. \ref{Fig:Evolution}, we consider $5 \theta_i =1$. In this case, after the Hubble friction term becomes small enough compared to the mass term, the pNGB goes to the minimum of the potential and subsequently crosses the barrier when the kinetic energy becomes equal to the potential barrier, i.e. $\frac{1}{2}\dot{\theta}^2(T) =  \frac{2m_{a}^2(T)}{25}$. It then starts to slide across the rapidly decreasing potential barriers (cf. Fig. \ref{Fig:schematic}). The left panel shows the evolution of the Hubble rate $H$, the temperature-dependent mass $m_a (T) \times \frac{2}{5}$, the velocity $|\dot{\theta}|$ and the comoving energy density of the oscillation $\rho_{\rm osc} /s$. The scaling of $|\dot{\theta}|$ (magenta contour) changes from $T$ to $T^3$ after $T_c$, and finally the pNGB starts to oscillate when the velocity becomes comparable to the mass $m_a^{(0)}$(blue contour). On the other hand, the oscillation energy density (black contour) keeps on diluting until it becomes constant after $T_{\rm osc}$. The right panel shows the evolution of $\theta$, which is frozen initially at the value $\theta_i$. It increases after the temperature reaches $T_{\rm slide}$, since it continues to cross the potential barriers as discussed above. The oscillation of the pNGB after $T_{\rm osc}$, can be seen clearly in the magnified inset plot. The different dashed lines in the figures indicate our analytical estimates of $T_{0}$, $T_{\rm slide}$, $T_{\rm osc}$ and $\rho_{\rm osc} /s$, given by Eq. \eqref{eq:T0}, \eqref{eq:Tslide}, \eqref{Eq:Tosc} and \eqref{Eq:DM} respectively, and we consider $T_c = f_a^{0}/\sqrt{c_\lambda}$. Starting from a reheating temperature $T_{\rm RH}\simeq 10 ~T_0$, we find that $\mathcal{C}\simeq 3$ in the determination of $T_{\rm slide}$ (cf. Eq. \eqref{eq:Tslide}) provides a good estimate in order to match with our numerical results.

In Fig. \ref{Fig:Evolution2}, we show the behavior for a smaller value of the initial angle, considering $5 \theta_i =0.5$. Consequently, our pNGB cannot cross the barrier after reaching the minimum from the initial displaced value. Rather, it oscillates until the kinetic energy equals the potential barrier, after which it starts to slide, following the same behavior as in Fig. \ref{Fig:Evolution}, but with a relatively smaller oscillation energy density. Note that our predictions of the physical behavior of the pNGB, including its energy density given by Eq. \eqref{Eq:DM}, match adequately with our numerical solution.  


Finally, to verify our results further, we show the behavior of $|\dot{\theta}|/T$ for a wide range of our parameters and compare it with our analytical formula (cf. Eq. \eqref{Eq:thdotmax}) at $T_{\rm slide}$. For instance, we vary the pNGB mass $m_a^{(0)}$ from eV to the GeV scale. The results are shown in Fig. \ref{Fig:thetadot}, for $5 \theta_i =1$ (left panel) and $5 \theta_i =0.5$ (right panel). The solid lines indicate the evolution of the $|\dot{\theta}|/T$ obtained numerically for different choices of benchmark values covering a wide range. The dashed horizontal line represents the analytical estimate, given by Eq. \eqref{Eq:thdotmax}. We again find that ${\mathcal C}\simeq3$, considering $T_{\rm RH}\simeq 10 ~T_0$. The vertical dashed lines represent our analytical estimates of the temperatures at the boundaries of the different regimes, which again, we find to agree reasonably well with our numerical results.

\section{Growth of \lowercase{p}NGB fluctuations}\label{ApC}

In this section, we explore the evolution of the fluctuations of the pNGB field. The growth of these fluctuations might lead to a loss of energy in the homogeneous mode, eventually stopping the motion of the pNGB. Such a phenomenon known as axion fragmentation was first explored in Ref.~\cite{Greene:1998pb} when the pNGB oscillates around the potential minimum, finding that the fragmentation is important for large misalignment angle that can be motivated by some inflation dynamics~\cite{Co:2018mho} or anthropic argument~\cite{Freivogel:2008qc}. The application on structure formation enhancement is further studied in Ref.~\cite{Arvanitaki:2019rax}. Refs. \cite{Fonseca:2019lmc, Jaeckel:2016qjp, Berges:2019dgr} studied fragmentation in the context of relaxion and axion monodromy models. Fragmentation in more general situations including pNGB oscillating around the minimum and sliding over multiple potential barriers is comprehensively investigated in Refs. \cite{Fonseca:2019ypl,Eroncel:2022vjg, Eroncel:2022efc}. 
In both cases, parametric resonance~\cite{Kofman:1994rk,Kofman:1997yn} can play an essential role in the growth of pNGB fluctuations if certain conditions are satisfied, e.g. the ratio between the pNGB mass and the Hubble parameter being large enough. 
As will be shown in the following, the pNGB in our scenario also experiences sliding and oscillating stages, and
can experience significant fragmentation. 
Therefore, we use semi-analytical method, following Refs.~\cite{Fonseca:2019ypl, Eroncel:2022vjg}, and numerical calculation to investigate the effects of fragmentation on our main results.

We start by decomposing the pNGB field into the homogeneous mode and the fluctuation part as
\begin{align}
    \Theta(t,\textbf{x})= \theta(t) + \delta \theta (t,\textbf{x})\, ,
\end{align}
which is valid for $ |\delta \theta (t,\textbf{x})/\theta(t)| \ll 1 $. By definition, this means that fragmentation is not sufficient so we can still distinguish the background and the fluctuations. 
This condition is violated when the energy density of the produced fluctuations is comparable to that of the homogeneous mode, which implies that the backreaction from the fragmentation is non-negligible. 
The following analysis is based on the validity of this decomposition, so it breaks down at the point of significant fragmentation which should be studied with sophisticated numerical methods such as lattice simulation~\cite{Eroncel:2022vjg, Eroncel:2022efc}. 
Fortunately, for our purpose, it is sufficient to just identify the occurrence of strong fragmentation, which can be done by finding out whether the energy densities of the homogeneous mode and that of the fluctuations become comparable, 
without considering the evolution after backreaction becoming important. 

The fluctuation $\delta \theta (t,\textbf{x})$ is decomposed into the Fourier modes as 
\begin{align}
    \delta \theta (t,\textbf{x})= \int \frac{d^3k}{(2\pi)^3}\delta\theta_\textbf{k} (t) e^{-i \textbf{k.x}}\,,
\end{align}
where $\delta\theta_\textbf{k} (t) =\delta\theta_k \hat{\delta\theta_\textbf{k}}$ with $\langle\hat{\delta\theta_\textbf{k}}\hat{\delta\theta_{\textbf{k}^{\prime}}^*} \rangle= (2\pi)^3\delta^{(3)}(\textbf{k}-\textbf{k}^\prime)$. 
The equation of motion for $\delta\theta_k$ is given by 
\begin{align}
\delta\ddot{\theta}_k+\left(3H+2\frac{\dot{f}_a}{f_a}\right)\delta\dot{\theta}_k+\left[\frac{k^2}{R^2}+m_a^2(T)\cos {\left( n \theta \right)} \right]\delta\theta_k=0\,,\label{eq:eomthetk}
\end{align}
where $R(t)$ denotes the scale factor in the flat FLRW metric and $n=5$ in our case. 

As previously mentioned, the evolution of the pNGB in our scenario experiences both sliding and oscillating phases.  
To unify the discussion on these stages, it is convenient to define the dimensionless parameter $\varrho$ which is the ratio between the total pNGB energy density $ \rho_{\theta} $ and the height of its potential barrier as 
\begin{align}
    \varrho (t) \equiv \frac{\rho_{\theta}}{2 m_a^2 f_a^2/n^2} = \frac{n^2}{4} \frac{\dot{\theta}^2}{m_a^2} + \sin^2 \left( \frac{n}{2} \theta \right) ~.
\end{align}
By definition, $\varrho>1$ means the pNGB can overcome the potential barrier, which indicates the pNGB sliding phase, whereas, for $\varrho<1$, the pNGB gets trapped and oscillates around the potential minimum. At the time of trapping at $T_{\rm osc}$, we have $\varrho (T_{\rm osc}) \simeq 1$. For small $ n\theta \ll 1 $ (i.e. $ \varrho <1 $) and when the Hubble expansion is negligible, Eq.\,\eqref{eq:eomthetk} can be approximated as the Mathieu equation, so $ \delta \theta_k $ may experience parametric resonance if $k$ falls within the instability band (or resonance band). 
 For more general cases such as large $ n\theta $ or the sliding phase ($\varrho>1$) we are interested in, the Mathieu equation approximation easily breaks down, so we need to employ the techniques developed in Refs.~\cite{Greene:1998pb,Eroncel:2022vjg}. Since we are mostly interested in the fragmentation effect on the sliding pNGB, we will implicitly assume $\varrho>1$ in the following. 

It is found that parametric resonance can also happen during the sliding phase for which the instability band is given by~\cite{Eroncel:2022vjg} 
\begin{align}
\varrho-1\leq \kappa^2\leq \varrho \,,\label{eq:insblbnd1}
\end{align}
with the dimensionless physical momentum $\kappa \equiv k/(m_a R) $ and Hubble expansion neglected. 
The normalization of $R$ does not change the physics while it appears in the form of $k/R$ for a given $k$ mode.
Thus, it is useful to define the size of the physical momentum scale for a given reference point, normalized by the pNGB mass as
\begin{align}
   \kappa_*= \frac{k}{m_{a*} R_*} \,,
\end{align}
with `$*$' indicating quantities evaluated at 
a reference time which can be arbitrary.
In the following, we take this reference time to be the ``would-be" trapping temperature $T_{\rm osc}$ given in Eq.\,\eqref{Eq:Tosc} (the ``actual" trapping temperature will be shifted from $T_{\rm osc}$ because of the fragmentation as we discuss in this section).
Because the kinetic energy of $ \theta (t) $ is much larger than its potential energy, $ \theta(t) $ can cross multiple potential barriers, during which the velocity (or kinetic energy) rapidly oscillates with small amplitude (besides decreasing due to Hubble expansion in a more realistic setup). 
To remove the unimportant small oscillations and grasp the essence of the resonance band, we take a time average of the condition~\eqref{eq:insblbnd1} which becomes  
\begin{align}
    \frac{n^2 \langle \dot{\theta}^2 \rangle}{4m_a^2} - \frac{1}{2}\lesssim \kappa^2 \lesssim \frac{n^2\langle \dot{\theta}^2 \rangle}{4m_a^2} + \frac{1}{2}\,.\label{eq:insblbnd} 
\end{align}
It can be rewritten into a useful form by expanding the square root as 
\begin{align}
    \left|\kappa-\kappa_{\rm cr} \right|\lesssim  \delta \kappa_{\rm cr} ~,~\kappa_{\rm cr}\equiv \frac{n \langle \dot{\theta}^2 \rangle^{1/2}}{2m_a} ~,~ \delta \kappa_{\rm cr} \equiv \frac{m_a}{2n \langle \dot{\theta}^2 \rangle^{1/2}}\, ,\label{eq:insblbnd2} 
\end{align}
where $ \langle \rangle $ denotes the time average. 
For notation simplification, we will omit $ \langle \rangle $ in the following and the time average is automatically implied. 
When Hubble expansion is taken into account, the upper and lower edges of the band is time-dependent because of the temperature-dependence of $ m_a (T) $ and $ \dot{\theta}(T) $. 
However, this analysis remains a good approximation as long as $ m_a (T) $ and $ \dot{\theta}(T) $ vary with time adiabatically within the typical time scale of the system, in other words, $ H \ll m_a, |\dot{\theta}| $ which is satisfied in our case. 
As will be shown later, the analytical results based on this approximation are consistent with the numerical results where we directly solve Eq.\,\eqref{eq:eomthetk} without any approximation. 

Under the aforementioned conditions, the effective mass of $ \delta \theta_k $ given in Eq.\,\eqref{eq:eomthetk} is a periodic function (up to the small oscillations in $\dot{\theta}$), so the solution for $ \delta \theta_k $ can be given in the Floquet form where the dominant (or exponentially growing part) is written as 
\begin{align}
    \delta \theta_k \sim  \delta\theta_k^i A_k (t) N_k (t) ~,~ N_k(t) = e^{\mu_k t} ~,\label{eq:delthetkan}
\end{align}
where $ \delta \theta_k^i $ is the initial value for $ \delta \theta_k $, $ A_k(t) $ is a periodic function of time, and $ \mu_k $ is the Floquet exponent describing the possible enhancement due to parametric resonance if $ {\rm Re}\{\mu_k\} >0 $. 
We will focus on the case where $ \mu_k $ is real and positive because we are interested in the fragmentation regime. 
Before going into the calculation of $ \mu_k $, the discussions for the other two factors are in order. 
In the presence of (adiabatic) Hubble expansion, the amplitude of $ A_k(t) $ decreases in time with the damping factor $A_k(t)\propto \omega_k(t)^{-1/2} R^{-y/2}$ where $ \omega_k^2 \equiv k^2/R^2 + m_a^2 \cos (n\theta) $. 
The value of $y$ on the scaling in terms of $R$ depends on the effective form of the friction term, i.e. $ y H \delta \dot{\theta}_k $ in Eq.\,\eqref{eq:eomthetk} at different stages of evolution. Interestingly, during the initial sliding, i.e. $y=1$, $A_k\propto \omega_k^{-1/2}R^{-1/2}\sim (k/R)^{-1/2}R^{-1/2}$ remains constant, which is consistent with our numerical results as we will shortly see.
On the other hand, the determination of the initial condition $ \delta \theta_k^i $ is rather complicated because it is affected by the curvature perturbation generated during inflation as a source term. 
Assuming any isocurvature perturbations are negligible, we simply take $ \delta \theta_k = 0,~ \delta \dot{\theta}_k =0 $ as the initial conditions in our numerical calculation. 
These will not result in trivial solutions because, on the right-hand side of the equation of motion for $ \delta \theta_k $, we have the source term, with the dominant contribution given by~\cite{Eroncel:2022vjg,Eroncel:2022efc}
\begin{align}
    S_k = 2\Phi_k m_a^2 \sin (n\theta) -4 \dot{\Phi}_k \dot{\theta} ~,
\end{align}
where $ \Phi_k $ is the Fourier mode of the curvature perturbation in the Newtonian gauge. 
During the radiation-dominated epoch, it evolves as 
\begin{align}
    \Phi_k (t) = 3\Phi_k (0) \left( \frac{\sin t_k - t_k \cos t_k}{t_k^3} \right)~,
\end{align}
where $ t_k = k/(\sqrt{3}RH) $. 
It is related to the comoving curvature perturbation as $ \Phi_k (0) =2 {\mathcal R}_k (0) /3 $ during inflation, so we can determine it by the CMB measurement of power spectrum of $ \mathcal{R}_k $ as 
\begin{align}
    \langle|\Phi_k (0)|^2\rangle&= \left(\frac{2}{3}\right)^2 \left(\frac{2\pi^2}{k^3}\right) \Delta_{\mathcal{R}}^2(k) \nonumber \\&= \left(\frac{2}{3}\right)^2\left(\frac{2\pi^2}{k^3}\right)A_s\left(\frac{k}{k_{\rm pivot}}\right)^{n_s-1} ~,
\end{align}
where $ A_s = 2.1\times 10^{-9} $ and $ n_s = 0.9649 $ at $ k_{\rm pivot}=0.05~{\rm Mpc}^{-1} $~\cite{Planck:2018vyg}. 
In our calculation, we simply take $ n_s =1 $. 
To compare the analytical results with the numerical ones, we set $ \delta \theta_k^i = \delta \theta_{k,{\rm num}} (t_{k, {\rm amp}}) $ for our analytical formula, where $ \delta \theta_{k,{\rm num}} (t_{k, {\rm amp}}) $ is the numerical solution for $ \delta \theta_k $ at the time $ t_{k, {\rm amp}} $ when the given $ k $ mode begins to be amplified due to parametric resonance. 

The last piece of our analytical calculation is the Floquet exponent. 
Generally, $ \mu_k $ is time-dependent, so the total enhancement should be integrated with respect to time $ N_k = \exp (\int \mu_k dt) $. 
But within each typical time scale of the system, i.e. the time for the background $ \theta $ to cross one potential barrier, $ \mu_k $ changes only adiabatically, so we can approximate it as a constant and the integration as $ \sim \mu_k \Delta t$ with $\Delta t$ being one period. 
Focusing on each period and $ \varrho \gg 1 $, the Floquet exponent is given by~\cite{Fonseca:2019ypl,Eroncel:2022vjg} 
\begin{align}
    \mu_k &\approx \left[ \delta \kappa_{\rm cr}^2 - \left( \kappa-\kappa_{\rm cr} \right)^2 \right]^{1/2} m_a ~\overset{\kappa=\kappa_{\rm cr}}{\simeq} \frac{m_a^2}{2n\dot{\theta}} \label{eq-muk}~,
\end{align}
where the second equation is for the most enhanced mode. 

Before going into detailed calculation for different stages of pNGB evolution, we quantitatively set the criterion for efficient fragmentation at which our calculation breaks down and the backreaction has to be taken into account. 
This is convenient for our subsequent discussion, although the full non-linear calculation is beyond the scope of this work.  We define the moment of efficient fragmentation as the time when the energy density of the produced fluctuations equals that of the homogeneous pNGB background. 
The energy density of the fluctuations is calculated as 
\begin{align}
    \rho_{\rm fl} (t) = \int \frac{d^3k}{(2 \pi)^3} \rho_{\delta\theta_k}\,,  \label{eq:rhofl1}
\end{align}
where for each $ k $ mode 
\begin{align}
   \rho_{\delta\theta_k}= \frac{f_a^2(t)}{2}\left[|\dot{\delta\theta_k}|^2+\left(\frac{k^2}{R^2}+m_a^2(T)\cos{(n \theta)}\right)|\delta\theta_k|^2\right].\label{eq:rhothetk}
\end{align}
The maximum contribution to the energy density comes from the modes around $ k_{\rm cr}$, so we can further estimate the total energy density approximately as 
\begin{align}
    \rho_{\rm fl} (t) \sim \frac{1}{2\pi^2} k_{\rm cr}^2 \Delta k_{\rm cr} \times \rho_{\delta \theta_k}(k_{\rm cr},t)  \,. 
    \label{eq:rhofl2}
\end{align}
where $\Delta k_{\rm cr}$ is the full-width at half-maximum (FWHM) of the spectrum of $\rho_{\delta \theta_k}$. The expression above will be useful later to demonstrate the evolution of the energy density of the fluctuations, as we will see shortly.

Once $ \rho_{\rm fl} = \rho_{\theta} $ is satisfied, the fragmentation due to parametric resonance is sufficient to stop the sliding. In the following, we will discuss the fragmentation in our scenario and take $ n=5 $. 
Specifically, there are two regimes where $m_a$ and $\dot \theta$ scale differently, depending on whether the temperature is above or below $T_c$, which can lead to different behaviors for the fluctuations.
Hence, we discuss these two cases separately below.

\begin{figure*}
    \centering
    \includegraphics[width=0.445\textwidth]{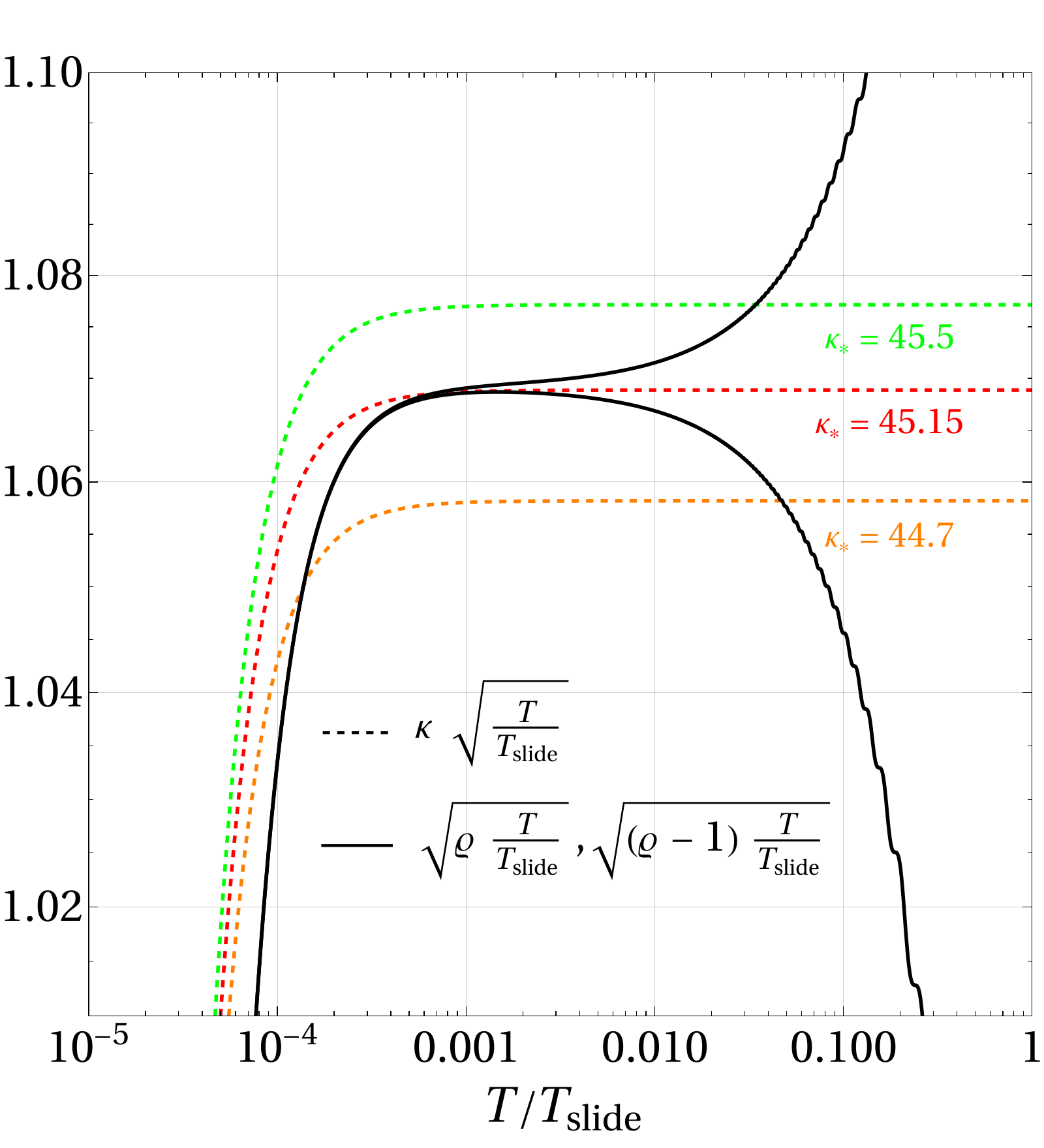}
    \includegraphics[width=0.48\textwidth]{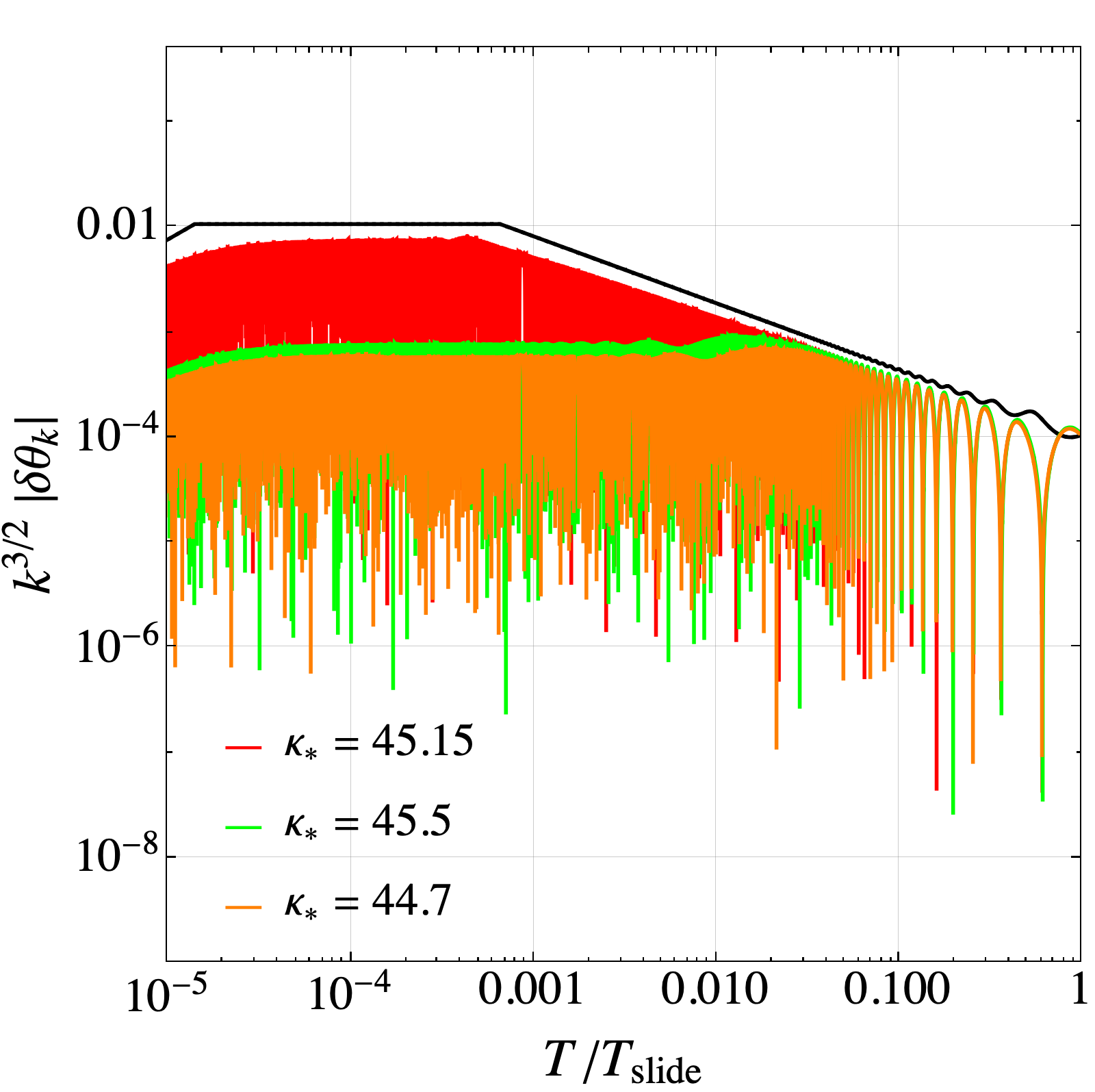}
    \caption{\textit{Left panel}: Evolution of the instability band with upper and lower boundaries given by $\sqrt{\varrho} $ and $ \sqrt{\varrho-1} $ during the initial sliding phase ($\dot{\theta}\propto T$) along with different modes $ \kappa $. For a better visualisation of this very narrow band, we multiply the wave numbers as well as the resonance bands by $ (T/T_{\rm slide})^{1/2}$, since $\kappa$ scales as $ (R\,m_a)^{-1}\propto T^{-1/2}$ (see text). \textit{Right panel}: Evolution of the fluctuation $\delta\theta_k$ considering the same values of $ \kappa_* $ as in the left panel plot. The black dashed line indicates our analytical result (Eq. \eqref{eq:Nkearly}) for the mode with maximum amplification.}
    \label{Fig:PR1}
\end{figure*}

\textbf{Initial sliding ($\dot{\theta}\propto T$, $m_a\propto T^{3/2}$) :}\\
During the initial phase of sliding, $\dot{\theta}\propto R^{-1}$ and $m_a \propto R^{-3/2}$, so $ \kappa/\kappa_{\rm cr} $ stays constant, or in other words, the physical momentum redshifts in approximately the same manner as the edge of the band. This implies that the most enhanced mode,  $ k=R_* m_{a*} \kappa_{\rm cr*} $, which corresponds to the center of the instability band can be ``trapped" inside the band as the Universe expands until $T=T_c$. However, the width of the band is very narrow since $\varrho\gg1$, and decreases following  $ \delta \kappa_{\rm cr} \propto R^{-1/2} $. Thus, the momentum mode escapes the band slightly before reaching $T=T_c$, when the scaling of $\dot\theta$ and $m_a$ starts to change slightly from $R^{-1}$ to $R^{-3}$, and $R^{-3/2}$ to constant, respectively. In the left panel of Fig. \ref{Fig:PR1}, we show the evolution of the resonance band during this initial sliding phase where $\dot \theta$ and $m_a$ are obtained numerically for the model parameters ($f_a^{(0)} = 10^{6} ~\text{GeV}$, $m_a^{(0)} = 1 ~ \text{eV}$, $c_{\lambda} = 10^6$) which correspond to those taken in Fig. \ref{Fig:Evolution} in the previous section.
Here, we take $5\theta_i =1$, but we will later discuss what happens when $5\theta_i$ is smaller.
The three colored dashed lines correspond to three examples of momentum modes whose $\kappa_*$ values are around $45.15$ which satisfies $ \kappa \simeq \kappa_{\rm cr} $. For a better visualisation of this very narrow band, we remove the scaling of $\kappa$, by multiplying all the relevant quantities, i.e. the wave numbers as well as the resonance bands by $ (T/T_{\rm slide})^{1/2}$, since $\kappa$ scales as $ (R\,m_a)^{-1}\propto T^{-1/2}$.

With Eq.~\eqref{eq-muk}, the growth of the modes at the central position of the band which amplify the most (for instance $ \kappa_*=45.15 $ in Fig. \ref{Fig:PR1}) is estimated as 
\begin{align}
    N_k &\simeq \exp{\left(\int_{t_s}^t \frac{m_a^2}{10\dot{\theta}} dt\right)} 
    \nn
    \\
    &\simeq \exp \left[ \frac{1}{2 \sqrt{\cal C}} \frac{1}{1-\cos(5\theta_i)} \ln \left(\frac{T_{\rm slide}}{T}\right) \right]\label{eq:Nkearly}
\end{align}
where we have used Eqs.~\eqref{eq:Tslide} and \eqref{Eq:thdotmax}. Note that the above slightly overestimates the growth since it is obtained by integrating over many periods over which $\dot{\theta}$ changes considerably, while the Floquet analysis ($N_k \sim \exp(\int\mu_k dt)$) holds for constant $\dot{\theta}$ or a small non-zero $\ddot{\theta}$.
Nonetheless, the result agrees substantially with the numerical results as shown in the right panel of Fig.~\ref{Fig:PR1} which exhibits the evolution of $\delta\theta_k$ (multiplied by $k^{3/2}$ to make it dimensionless), for different values of $\kappa_*$, chosen to be the same as in the left panel plot. 
The dashed black line shows our analytical ansatz for the mode with maximum amplification.  Comparing with the plot in the left panel, we can see that the modes are amplified until they leave the instability band. 

\begin{figure*}
    \centering
    \includegraphics[width=0.48\textwidth]{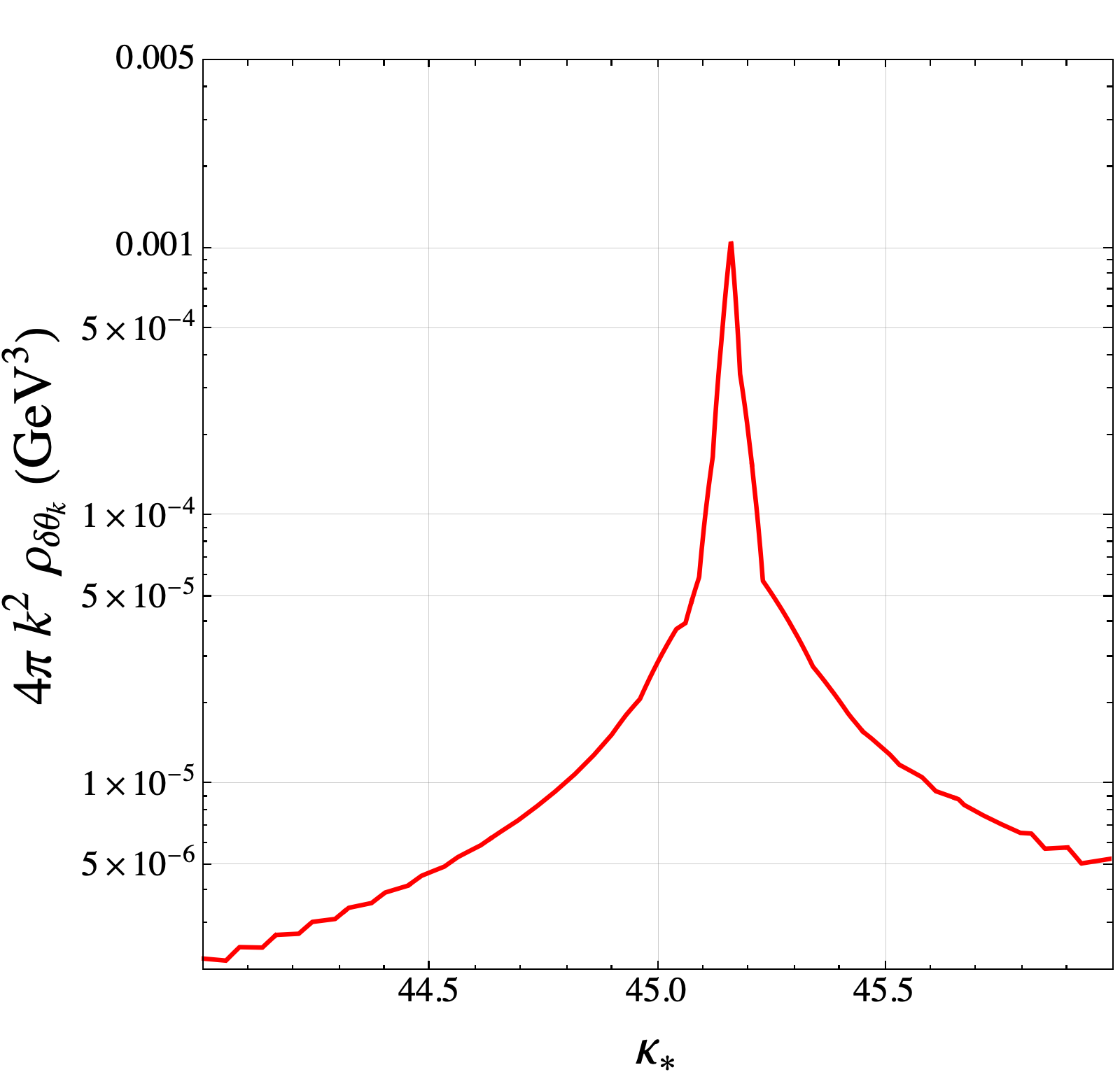}~~~
\includegraphics[width=0.46\textwidth]{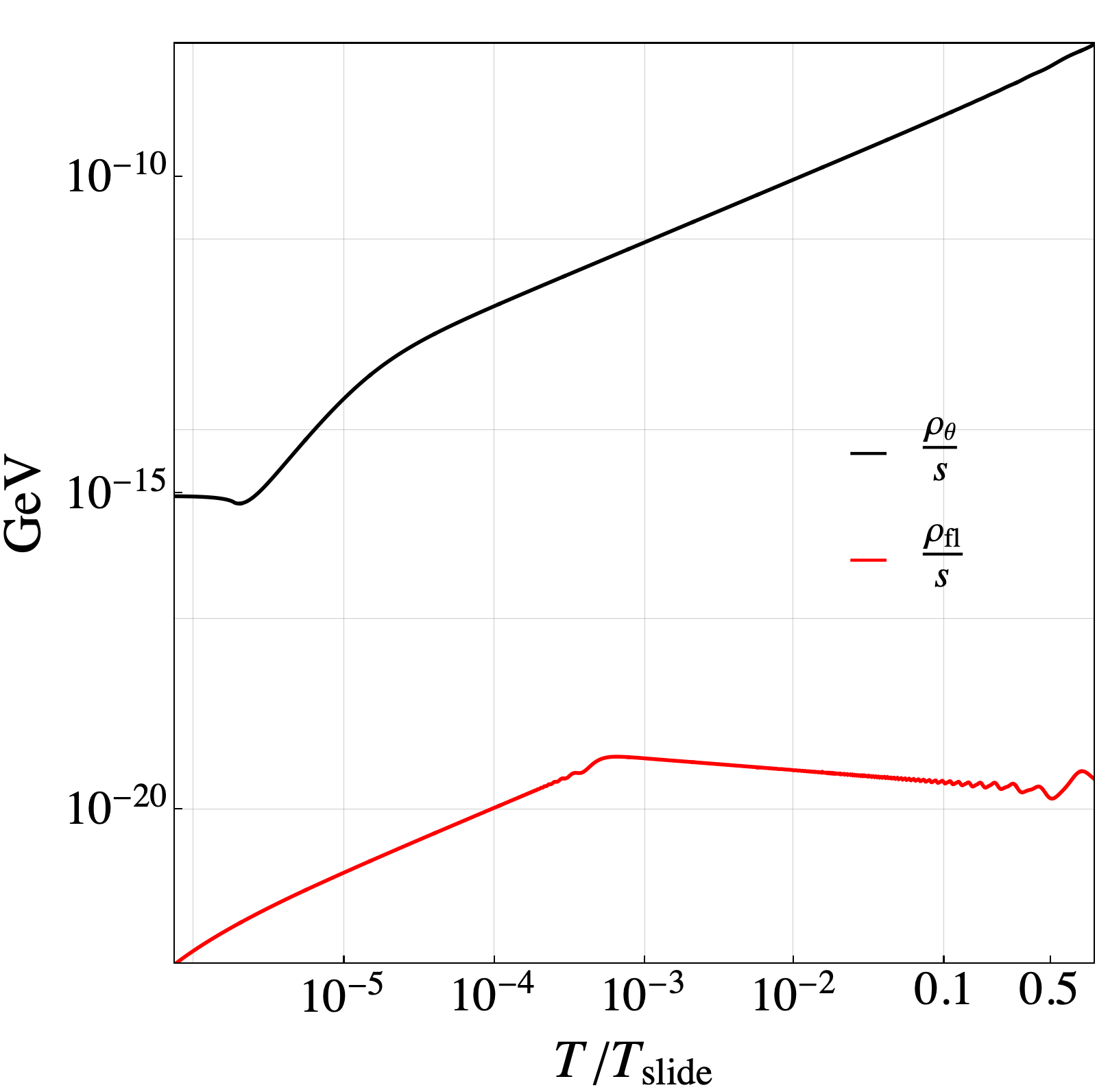}
    \caption{\textit{Left panel}: 
    The energy spectrum of the fluctuations at the end of amplification found numerically through Eq. \eqref{eq:rhothetk}. It is found to peak around $k_*\simeq 45.15$, in agreement with Fig. \ref{Fig:PR1}. \textit{Right panel}: Evolution of the energy density of the homogeneous mode and that of the fluctuations, the latter calculated using Eq. \eqref{eq:rhofl2}.}
    \label{Fig:rhok1}
    \end{figure*}

The growth during this stage is moderate in the sense that it is not enough to generate non-negligible backreaction to the background dynamics, which can be seen by comparing the energy densities $ \rho_{\rm fl} $ and $ \rho_{\theta} $.
In the left panel of Fig. \ref{Fig:rhok1}, we show the energy density spectrum of the produced fluctuations whose peak position agrees with the result in Fig.~\ref{Fig:PR1}. The evolution of the energy densities of both the homogeneous background and the fluctuations are shown in the right panel, with the latter calculated using our approximation given by Eq. \eqref{eq:rhofl2}, where we take the FWHM from the spectrum in the left panel. The total energy density of the fluctuations at the exit of the most amplified mode is found to be significantly smaller than that of the homogeneous mode,  $ \rho_{\rm fl} \sim \mathcal{O}(10^{-8}) \rho_{\theta} $. Using the full numerical results calculated from Eq.~\eqref{eq:rhofl1} by integrating over the relevant range of $\kappa_*$ shown in the left panel of Fig. \ref{Fig:rhok1} gives us $ \rho_{\rm fl} = 4.5 \times 10^{-8} \rho_{\theta}$ at the end of the amplification, which agrees with the approximation in Eq. \eqref{eq:rhofl2} up to a factor of less than $\mathcal{O}(1)$. Note that after the end of amplification, the energy density of fluctuations (cf. Eq. \eqref{eq:rhofl1}) redshifts as $\rho_{\rm fl}\propto f_a^2 R^{-2}\delta \theta_k^2 \sim R^{-4}$, while that of the background redshifts in the same manner, since $\rho_\theta \propto  f_a^2 \dot{\theta}^2 \sim R^{-4}$. However, below $T=T_c$, the background redshifts faster $\rho_\theta \propto   R^{-6}$, while the fluctuations still behave as radiation ($ \rho_{\rm fl}\propto  R^{-2} \omega_k^{-1} R^{-3} \sim R^{-4}$). Despite this, we find that the background energy density is not diluted enough to become comparable to that of the fluctuations.  

\begin{figure*}
    \centering
\includegraphics[width=0.48\textwidth]{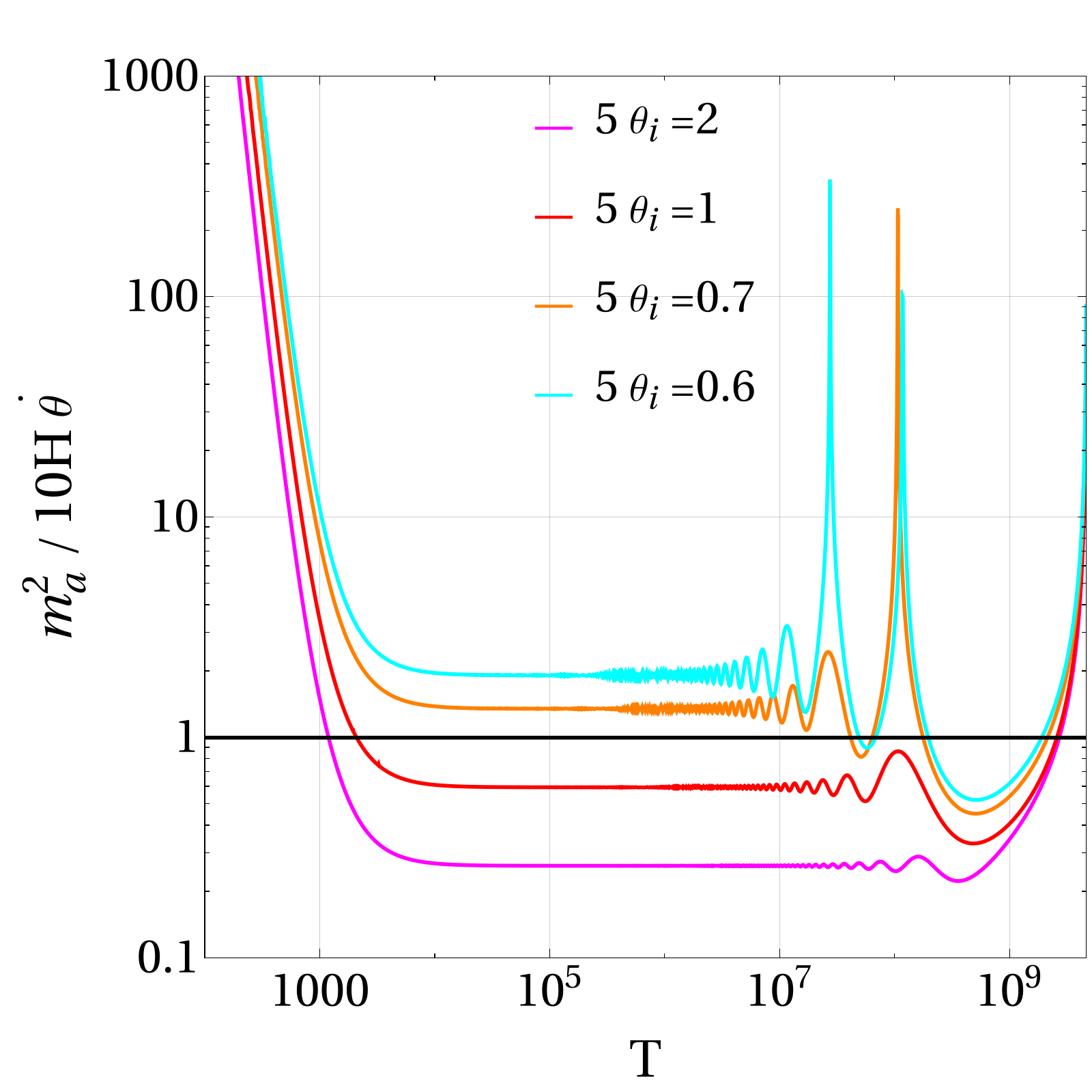}~~
\includegraphics[width=0.48\textwidth]{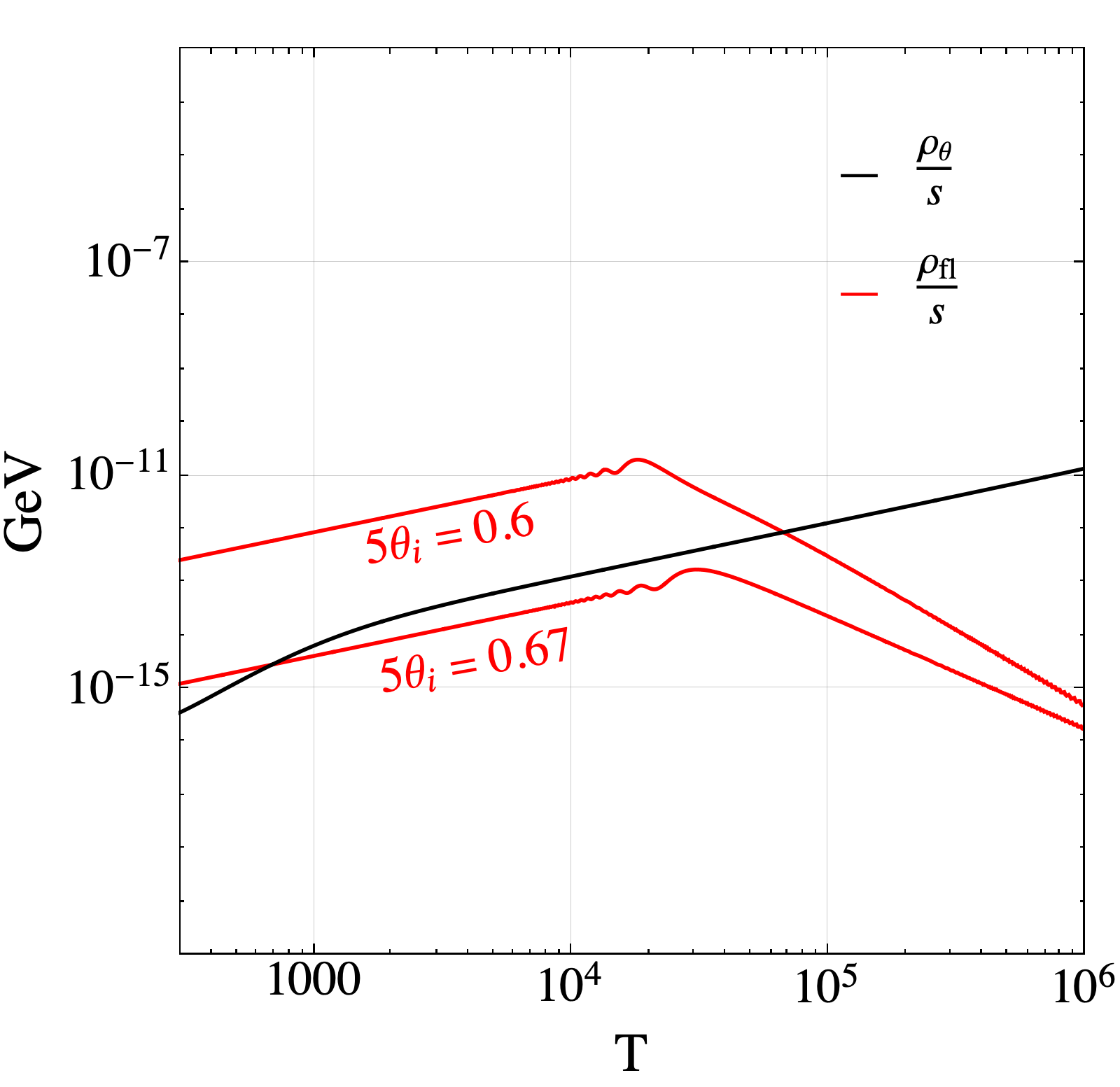}
    \caption{\textit{Left panel:} Variation of the ratio $ m_a^2/(10 \dot{\theta} H) $, considering different values of $5\theta_i$. During the initial sliding, it remains constant. The ratio becomes greater than unity once we approach the constant mass regime around $T=T_c$ and also for $5\theta_i\lesssim1$ (see text). \textit{Right panel:} Evolution of the energy density of the fluctuations for $5 \theta_i=0.6$ and $0.67$ (using Eq. \eqref{eq:rhofl2}) along with that of the background.}
    \label{Fig:comprH}
    \end{figure*}

Thus, for $ 5\theta_i \gsim \mathcal{O}(1)$ initial angle, we conclude that there is no significant enhancement of perturbations for the early stage of sliding, or no fragmentation. This is because the Hubble rate is not small enough compared to the growth rate,  which can seen easily from Eq.~\eqref{eq:Nkearly} by translating the integration over temperature into e-fold number $ -d\ln T =d N_e $. 
This means that the enhancement exponent is the product of $ N_e $ and $ m_a^2/(10 \dot{\theta} H) $ where the latter solely depends on $ \theta_i $ and $ \mathcal{C} $. For moderate $ N_e \sim \mathcal{O}(1) $ as in our case, the enhancement is important only when $ m_a^2/(10 \dot{\theta} H) \gsim 1 $ (similar condition in Ref. ~\cite{Fonseca:2019ypl}) which is not satisfied for $ 5\theta_i \gsim \mathcal{O}(1)$.  
In Fig. \ref{Fig:comprH}, we show the variation of this ratio with temperature, considering different values of the initial angle $\theta_i$. The constant values of this ratio agree considerably with our analytical estimate given as $(2\sqrt{\cal{C}}~(1-\cos5\theta_i))^{-1}$. For $5 \theta_i\gtrsim1$, the ratio is less than one, until around $T\simeq T_c$. Hence, the growth does not turn out to be significant as we saw from our numerical results earlier. However, for smaller values of $5\theta_i$ (for instance $5 \theta_i =0.7$), the Hubble rate becomes small enough during the amplification, such that this ratio gets enhanced. Calculating the energy density of the fluctuations in such a case, we find that it becomes comparable to that of the background. The right panel of Fig. \ref{Fig:comprH} shows the energy density evolution using our approximation given by Eq. \eqref{eq:rhofl2}, for two values of initial angle: $5\theta_i=0.6$ and $0.67$. While for the former, the energy density becomes equal to that of the background even before the end of amplification, the latter catches up the background energy density below $T=T_c$ because of the faster dilution of the background as discussed earlier. Thus, for our dynamics to remain unaffected, we restrict ourselves to $5\theta_i\gtrsim1$. 
In addition, after the mass saturates at $T_c$, the condition $ m_a^2/(10 \dot{\theta} H) \lesssim \mathcal{O}(1) $ is found to be  strictly violated, even for large initial misalignment angle $5\theta_i\gtrsim1$. This indicates that the enhancement after $ T =T_c $ can be significant even when the enhanced modes stay within the instability band for a short time, which we discuss below.

\textbf{Sliding during constant mass  ($\dot{\theta}\propto T^3$):}\\
In the calculation during this phase, $m_a=m_a^{(0)} ={\rm const} $. 
This phase is qualitatively the same as that studied in Refs.~\cite{Eroncel:2022vjg, Eroncel:2022efc} for the case of kinetic misalignment. 

\begin{figure}
    \centering
    \includegraphics[width=0.4\textwidth]{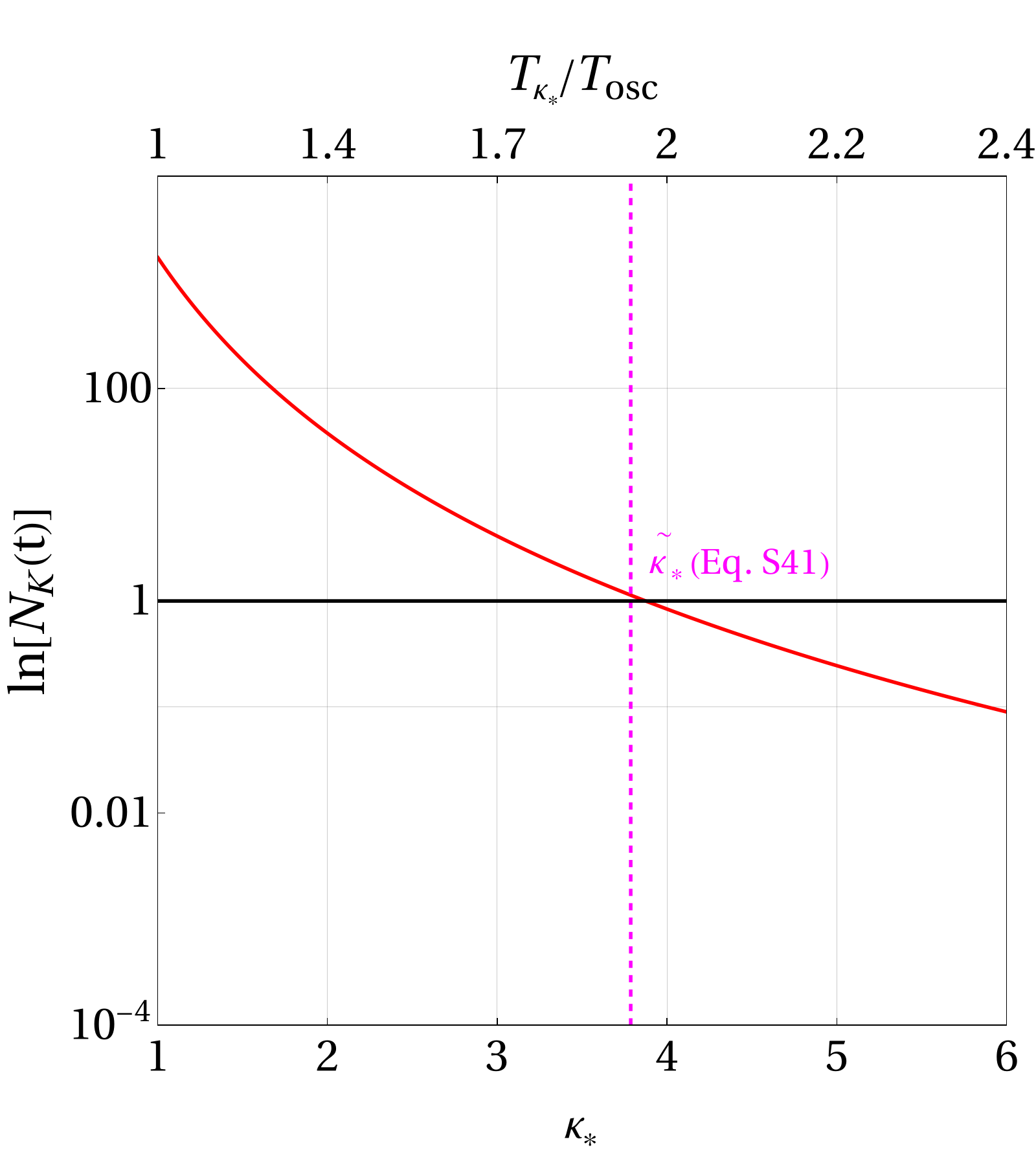}
    \caption{Variation of the amplification factor (Eq. \eqref{Eq:Nk_kappa_star}) with $\kappa_*= (T_{\kappa_*}/T_{\rm osc})^2$. It crosses unity at a temperature close to $ 2 T_{\rm osc}$, when $\kappa_* =\tilde{\kappa}_*$, where the magenta dashed line indicates our analytical estimate given in Eq. \eqref{eq:ktildest} or the square of Eq. \eqref{eq:TgrwTosc}.}
    \label{Fig:grw}
    \end{figure}

Note that in the earlier phase, there was a mode corresponding to the specific $\kappa_* \simeq \frac{5}{2}\frac{T_{\rm osc}}{m_a^{(0)}} \left( \frac{\dot\theta}{T} \right)_{\rm slide}$ which stays in the resonance band for a long time.
However, unlike the previous situation, each mode with $\kappa_*>1$ now has a short time period inside the resonance band.
Let us estimate this time period.
The temperature when the mode for a given $\kappa_*$ satisfies the critical resonance condition $\kappa=\kappa_{\rm cr}$
can be estimated as 
\bal
T_{\kappa_*} = \sqrt{\kappa_*} T_{\rm osc}.
\eal
This can be derived by solving $\kappa = \kappa_{\rm cr} \equiv \frac{5\dot \theta}{2m_a}$ (see Eq.\,\eqref{eq:insblbnd2}) with $\kappa \equiv \kappa_* \frac{m_{a*} R_*}{m_a(T) R(T)}=\kappa_* \frac{T}{T_{\rm osc}}$ and $\dot \theta (T)= \frac{2}{5} m_a^{(0)} \left(\frac{T}{T_{\rm osc}}\right)^3$.
The temperature $T_{\rm in}$ when the mode enters the resonance band can be obtained by solving $\kappa = \kappa_{\rm cr}- \delta \kappa_{\rm cr}$ (with $\delta \kappa_{\rm cr}=\frac{m_a}{10\dot\theta}=\frac{1}{4}\left( \frac{T_{\rm osc}}{T} \right)^3$ from Eq.\,\eqref{eq:insblbnd2}) and we find
\bal
H(T_{\kappa_*}) \Delta t = \frac{1}{8\kappa_* ^{3}} + O(\kappa_*^{-4}).
\eal
Then, the time scale defined by $2(\Delta t)$ gives us the time period of being inside the resonance band for a given mode defined by a given $\kappa_*$.
On the other hand, to evaluate $H(T_{\kappa_*})$, we need to trace back to our starting point $T_0$ when $m_a(T_0)=H(T_0)$, and re-express $T_{\rm osc}$ in terms of $T_0$. After that, we find 
\bal
\frac{m_a^{(0)}}{H(T_{\rm osc})} \simeq 
\left( \frac{{\cal C}^{1/2}}{2}(1-\cos 5\theta_i) \right)^{2/3}
\left(\frac{T_0}{T_c}\right)^{5/6},
\eal
and $H(T_{\kappa_*}) = \kappa_* H(T_{\rm osc})$.
Finally, since $\mu_k = \frac{m_a^2}{2 n \dot \theta} = \frac{{m_a^{(0)}}^2}{10\cdot (\frac{2}{5}{m_a^{(0)}} \kappa_*^{3/2})} = \frac{{m_a^{(0)}}}{4 \kappa_*^{3/2}}$, the amplification factor can be written as
\bal
N_k = \exp(\mu_k \times 2 |\Delta t|) = \exp \left[ \frac{m_a^{(0)}}{16 H(t_{\kappa_*}) \kappa_*^{9/2}} \right]
\nonumber \\ =
\exp \left[
\frac{
\left( \frac{{\cal C}^{1/2}}{2}(1-\cos 5\theta_i) \right)^{2/3}
}{16  \kappa_*^{11/2}}
\left(\frac{T_0}{T_c}\right)^{5/6}
\right].
\label{Eq:Nk_kappa_star}
\eal
Note that the subscript $k$ indicates that this is the amplification factor for a given $k$ mode, which has a one-to-one mapping with $\kappa_*$.
We define its critical value $\tilde \kappa_*$ by the condition $\log N_k = 1$, and obtain
\bal
\tilde \kappa_* \simeq 
\frac{
\left( \frac{{\cal C}^{1/2}}{2}(1-\cos 5\theta_i) \right)^{4/33}
}{16^{2/11}} \left( \frac{T_0}{T_c}\right)^{5/33}.\label{eq:ktildest}
\eal
As shown in Eq.\,\eqref{Eq:Nk_kappa_star}, the exponent has a large dependence on $\kappa_*$, and $N_k$ becomes exponentially huge as soon as $\kappa_*< \tilde \kappa_*$, i.e. $N_k = \exp[(\tilde \kappa_*/\kappa_*)^{11/2}]$.
Therefore, the criterion for having a complete fragmentation is given by $\frac{\tilde \kappa_*}{\kappa_*}\gtrsim1$, which depends on our parameter choices. In Fig. \ref{Fig:grw}, we show the variation of this amplification factor with $\kappa_*$, which can be related to the temperature as   $\kappa_*= (T_{\kappa_*}/T_{\rm osc})^2$. As we can see, it reaches unity around $\tilde{\kappa}_*$ (given by Eq. \eqref{eq:ktildest}), which is shown by the magenta-dashed line. 

This implies that as soon as the mode corresponding to $\tilde \kappa_*$ satisfies the resonance condition, the homogeneous mode gets fragmented, and the fluctuation with $\kappa_* \lsim \tilde \kappa_*$ dominates the total pNGB energy density.
The temperature $T_{\rm grw}$ when this happens can be simply estimated by $T_{\tilde \kappa_*} = \sqrt{\tilde \kappa_*} T_{\rm osc}$, and it can be written as
\bal
\frac{T_{\rm grw}}{T_{\rm osc}} &\simeq
\frac{
\left( \frac{{\cal C}^{1/2}}{2}(1-\cos 5\theta_i) \right)^{2/33}
}{16^{1/11}} \left( \frac{T_0}{T_c}\right)^{5/66}.
\eal
For a numerical check, let us replace $T_0$ by using its definition from the main text;
\bal
\frac{T_0}{T_c} \simeq
3 c_\lambda^2 \frac{{m_a^{(0)}}^2 M_{\rm Pl}^2}{g_* {f_a^{(0)}}^4}.
\eal
Then, we obtain the ratio $T_{\rm grw}/T_{osc}$ as
\begin{widetext}    
\bal
&\frac{T_{\rm grw}}{T_{\rm osc}} \simeq 4 \, \mathcal{C}^{1/33} \left[ 1-\cos 5\theta_i \right]^{2/33} \left( \frac{g_*}{100}\right)^{-5/66}
\left(\frac{c_\lambda}{10^8}\right)^{5/33} \left(\frac{m_a^{(0)}}{\text{eV}}\right)^{5/33}\left(\frac{f_a^{(0)}}{10^6\text{GeV}} \right)^{-10/33} ~,
    \label{eq:TgrwTosc}
\eal
\end{widetext} 
while $\tilde \kappa_*$ can be estimated by the square of it, i.e. $\tilde \kappa_* = (T_{\rm grw}/T_{\rm osc})^2$.
This indicates that in our parameter space given by Eqs.\,\eqref{Eq:ma0}, and \eqref{Eq:fa0}, the fragmentation occurs slightly before the would-be oscillation temperature $T_{\rm osc}$.
Thus, we can expect the effect to be small at this stage.

\begin{figure*}
    \centering
    \includegraphics[width=0.45\textwidth]{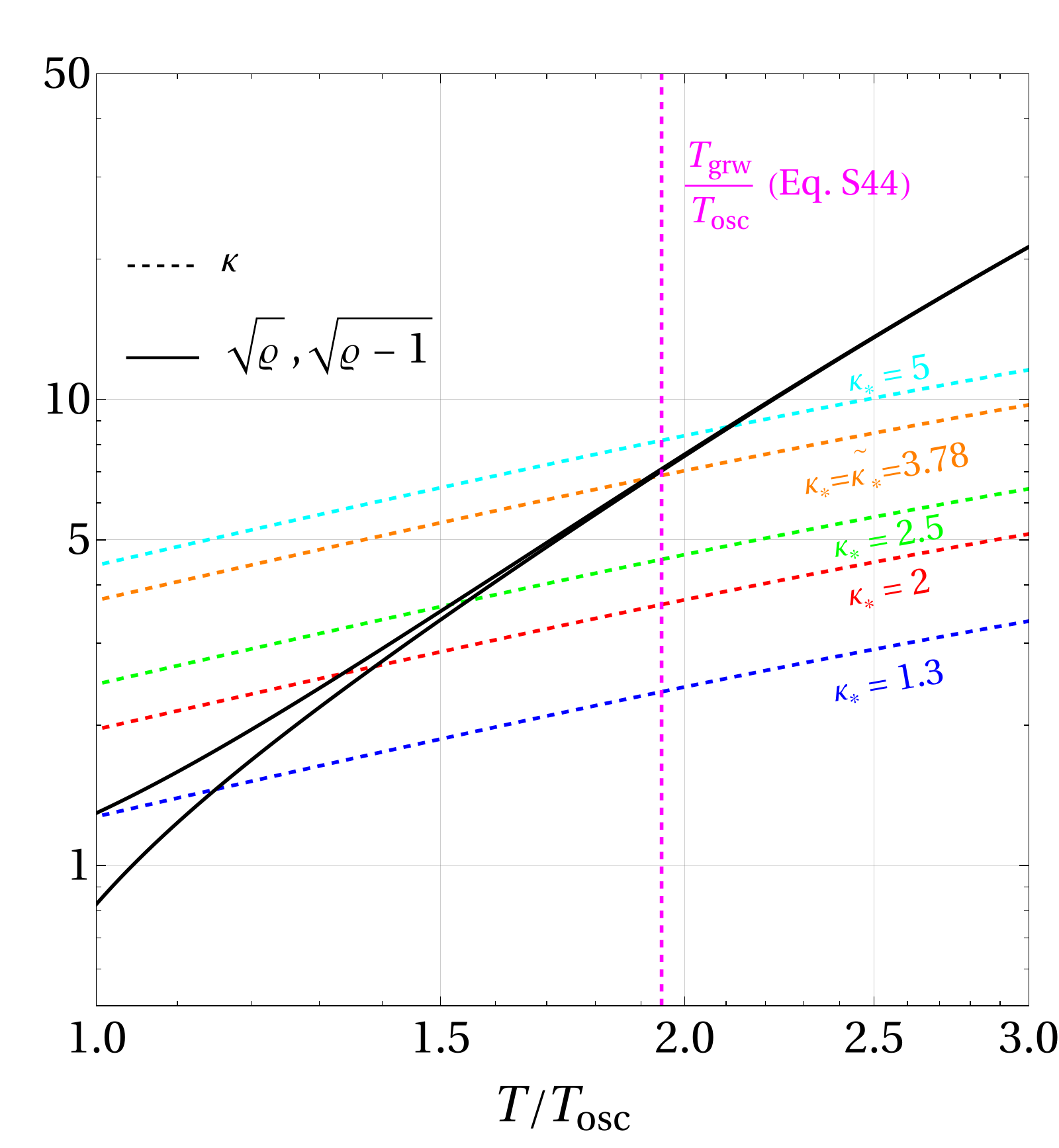}
    \includegraphics[width=0.48\textwidth]{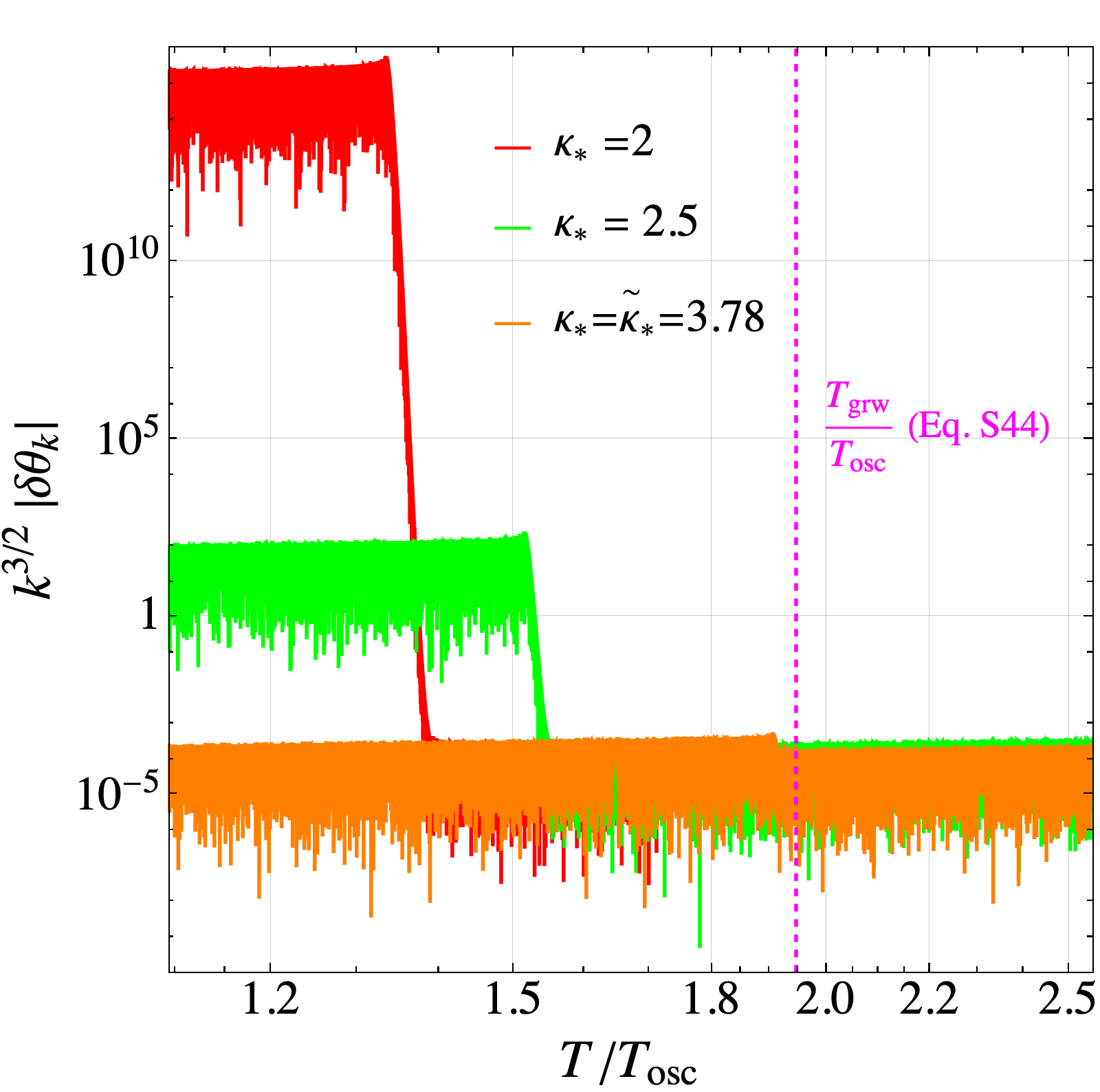}
    \caption{\textit{Left panel:} Evolution of the instability band, given by $\sqrt{\varrho}, \sqrt{\varrho-1} $ (Eq. \eqref{eq:insblbnd2}) during the sliding phase after $T=T_c$ ($\dot{\theta}\propto T^3$) along with the dimensionless wave number $\kappa$, considering some benchmark values of $\kappa_*$. \textit{Right panel}: Evolution of the fluctuations $\delta\theta_k$ for some $\kappa_*$ values of the left panel plot. As demonstrated in Fig. \ref{Fig:grw}, and also by our estimate in Eq. \eqref{eq:TgrwTosc}, the moment when the modes start to grow is shown by the vertical dashed line, which is in good agreement with our numerical results.}
    \label{Fig:PR2}
    \end{figure*}

Now, let us check our above analytical estimates with our numerical results.
In Fig. \ref{Fig:PR2}, we show the 
behavior of the modes corresponding to some benchmark values of $\kappa_*$. 
In the left panel, we show the evolution of $\kappa$ along with the resonance band depicted by the black solid line. As expected from Eq.\,\eqref{eq:insblbnd2} with constant $m_a$, we can see that the band widens over time.  Here, we choose the same benchmark parameters for the homogeneous mode as earlier. 
For these parameters, the moment when the amplification factor becomes larger than unity is shown by the vertical dashed line, with corresponding $\tilde{\kappa}_*=3.78$. The plot in the right panel shows the evolution of the fluctuations, for some values of $\kappa_*$. As we expect, the growth of the fluctuations becomes large when the mode spends a sufficiently long time within the resonance band. We can see that the modes with $\kappa_*\lesssim\tilde{\kappa}_*$  indeed start to grow exponentially. Our estimate of this growth temperature (vertical dashed line), given by Eq. \eqref{eq:TgrwTosc}, is found to agree quite well with our numerical results. 

\begin{figure*}
    \centering
    \includegraphics[width=0.48\textwidth]{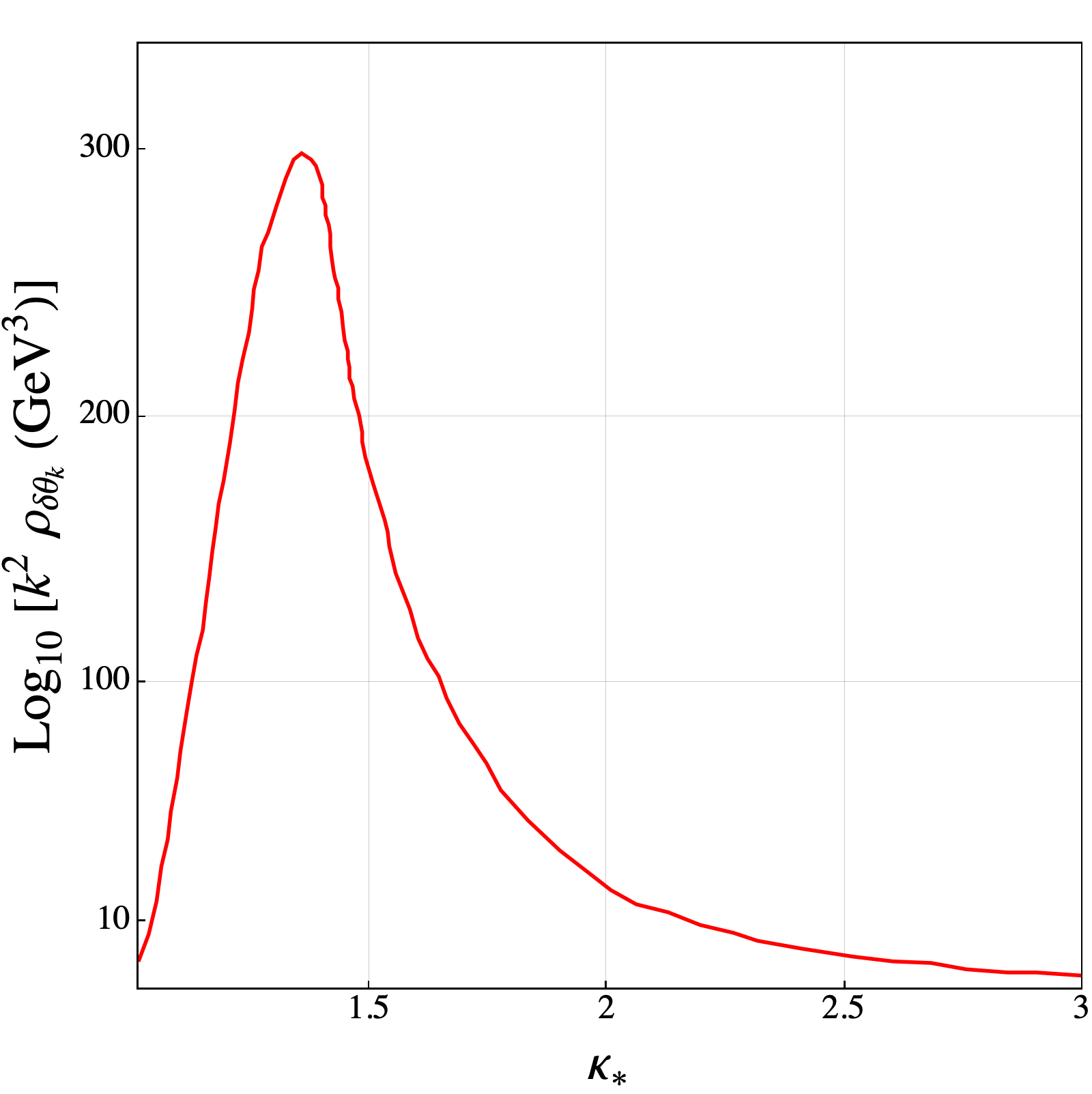}
    \includegraphics[width=0.48\textwidth]{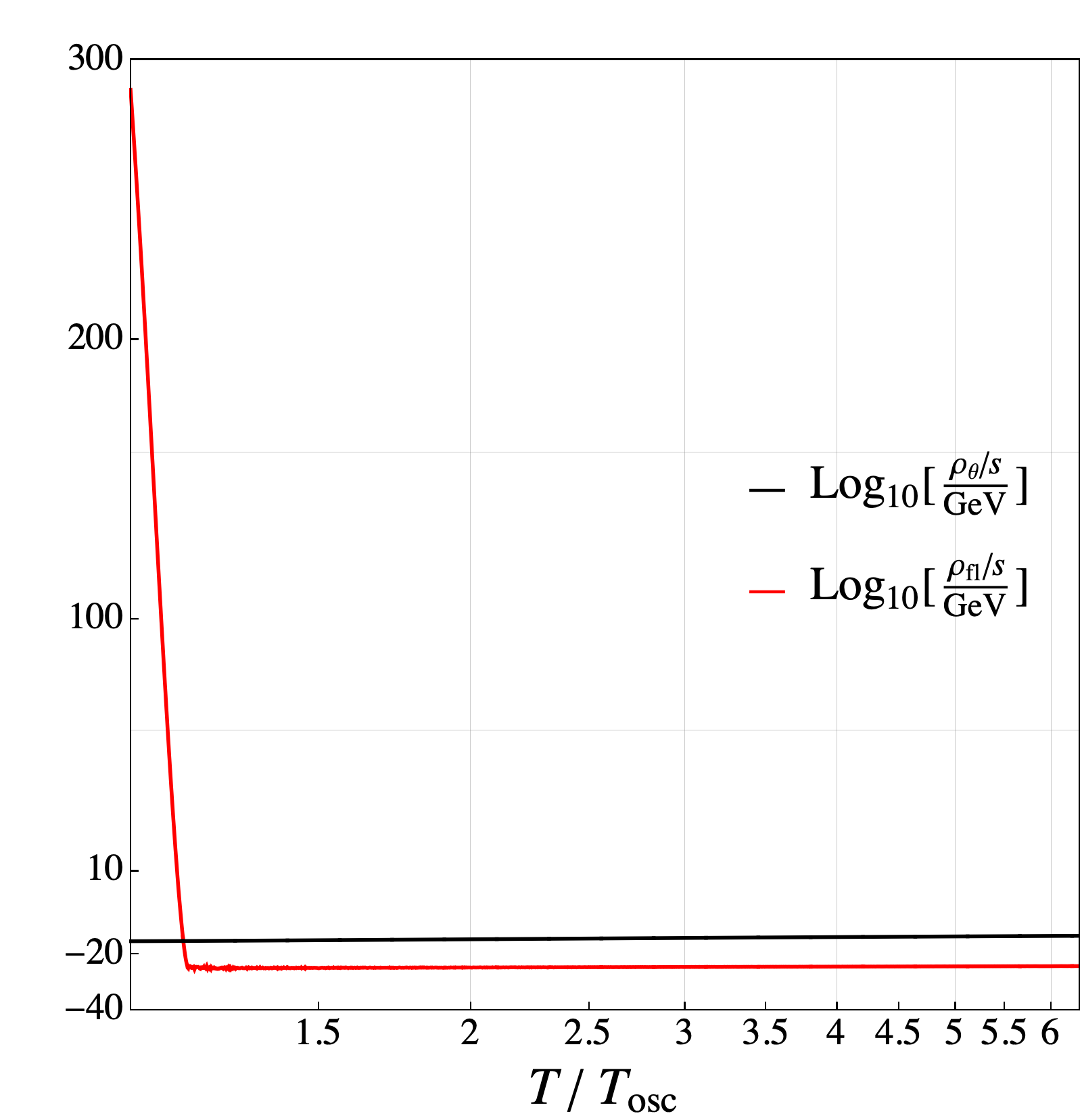}
    \caption{ 
    \textit{Left panel}: 
    The energy spectrum of the fluctuations at $T\simeq T_{\rm osc}$ found numerically through Eq. \eqref{eq:rhothetk}. It is found to peak at $\kappa_* = 1.36$, as also anticipated from the left panel of Fig. \ref{Fig:PR2}. \textit{Right panel}: Evolution of the energy density of the homogeneous mode and that of the fluctuations, the latter calculated using Eq. \eqref{eq:rhofl2}. The huge exponential growth indicates breakdown of our analysis at $T\lesssim T_{\rm grw}$.}
    \label{Fig:rhok2}
\end{figure*}

Thus, because of this huge exponential growth, the energy of the fluctuations is expected to increase exponentially, soon after the temperature reaches $T_{\rm grw}$ (or equivalently $\kappa_*\lesssim \tilde \kappa_*$), leading to fragmentation of the homogeneous mode at a temperature $T_{\rm frag}\simeq T_{\rm grw}$. Note that, considering $T> T_{\rm osc}$, the time spent inside the resonance band is maximized for modes entering the band close to $T_{\rm osc}$ (see for instance $\kappa_*=1.3$ in the left panel of Fig. \ref{Fig:PR2}). In the left panel of Fig. \ref{Fig:rhok2}, we show the energy density spectrum at $T\simeq T_{\rm osc}$, which is found to peak at $\kappa_*= 1.36$.  In the right panel, we show the evolution of the energy density until $T\simeq T_{\rm osc}$, using the semi-analytic approximation given by Eq. \eqref{eq:rhofl2}, where we only take the contribution from the maximally growing mode and consider the FWHM estimated numerically from the spectrum in the left panel. Clearly, because of the exponential growth, the energy is found to increase enormously.  From our full numerical results calculated by integrating Eq. \eqref{eq:rhofl1} over the relevant k range, we find $\rho_{\rm fl}= 4.5 \times 10^{304} ~\rho_{\theta}$ at $T\simeq T_{\rm osc}$ which again, agrees considerably well with our approximation in Eq. \eqref{eq:rhofl2}, upto a factor of less than $\mathcal{O}(1)$. Undoubtedly, our analysis breaks down much before this point, once the fluctuation energy density dominates over that of the background, which takes place just after $T_{\rm grw}$ because of the exponential growth (see Fig. \ref{Fig:PR2}), as discussed above. Hence, we take the temperature at the moment of fragmentation as $T_{\rm frag}= T_{\rm grw}$. Note that the fragmentation which occurs close to the \textit{would-be} oscillation temperature, does not affect our generated baryon asymmetry, which for our working parameter space, gets frozen during sliding, at temperatures higher than $T_{\rm frag}$.  Below, we estimate the possible modification to the relic density of pNGB.

\textbf{pNGB relic density modified by the fragmentation:}\\
In the presence of complete fragmentation, the energy density of the pNGB can get modified because the fluctuation is relativistic as $\kappa_*>1$ obviously indicates.
Thus, the energy density of the fluctuation scales as $\rho_{\rm fl}\propto R^{-4}$ after fragmentation, until it becomes non-relativistic. 
On the other hand, the homogeneous mode redshifts as $\rho_{\theta}\propto R^{-6}$ in our estimation without considering the fragmentation. 
Thus, whether the fragmentation actually enhances or decreases the relic density is determined by how large $\tilde \kappa_*$ is (assuming the dominant mode is not too far from $\tilde \kappa_*$).

We define the total ``modification" factor as
\begin{align}
    \mathcal{Z} \equiv \frac{\rho_{\rm fl}}{\rho_{\rm osc}}\, ,
\end{align}
where $\rho_{\rm fl}$ and $\rho_{\rm osc}$ denote the energy densities of the pNGB estimated with and without considering the fragmentation effects.
The fluctuation becomes non-relativistic at a temperature, say $T_{\rm nr}$, which can be estimated as 
\begin{align}
    \frac{k} {R}=m_a
    \quad \Rightarrow \quad 
    \frac{\tilde \kappa_* T_{\rm nr}}{T_{\rm osc}}=1 
    \quad \Rightarrow \quad 
    T_{\rm nr}= \frac{T_{\rm osc}}{\tilde \kappa_*}\,.\label{eq:Tnr}
\end{align}
Note that we have $\tilde \kappa_*>1$ from Eq.\,\eqref{eq:TgrwTosc}, and hence $T_{\rm nr}$ is lower than the \emph{would-be} oscillation temperature $T_{\rm osc}$ without fragmentation. 
\begin{figure}[htb!]
    \centering   \includegraphics[width=0.4\textwidth]{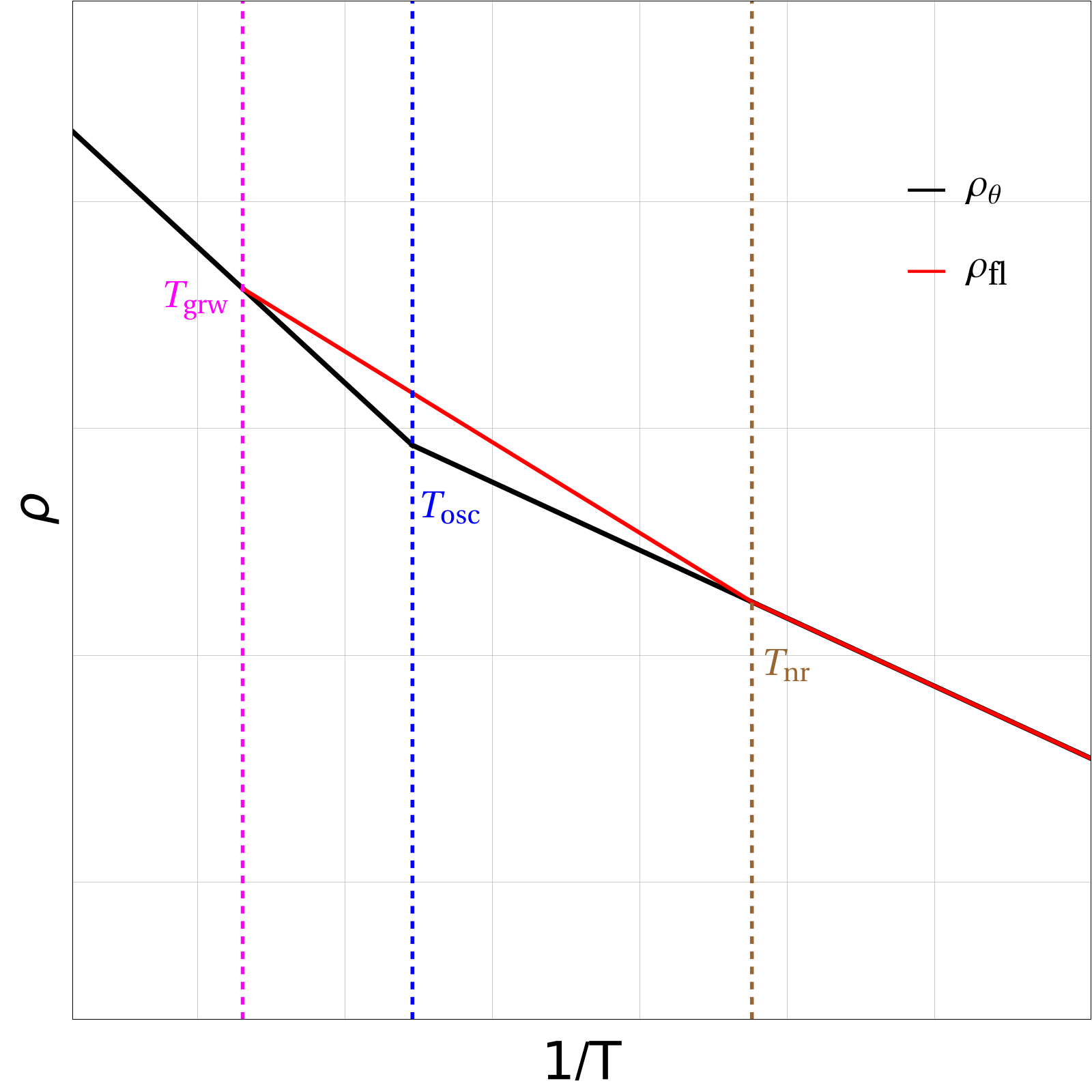}
    \caption{Schematic showing the energy density evolution of the homogeneous pNGB and the fluctuations. Note that $\rho_{\rm fl} (T_{\rm nr})$ turns out to be equal to  $\rho_{\theta} (T_{\rm nr})$ (see Eq. \eqref{eq:rhofluc0}) and thus leading to unchanged relic abundance.}
    \label{Fig:rlcschm}
    \end{figure}
The energy density of the fluctuations at any temperature below $T_{\rm nr}$, say $T_f$, is given by 
\begin{align}
    \rho_{\rm fl}(T_f)&\simeq \rho_{\rm fl} (T_{\rm nr})\left(\frac{T_f}{T_{\rm nr}}\right)^3\nonumber
    \\
    &\simeq \rho_{\rm fl} (T_{\rm grw})\left(\frac{T_{\rm nr}}{T_{\rm grw}}\right)^4\left(\frac{T_f}{T_{\rm nr}}\right)^3\nonumber
    \\
    &\simeq 
    \rho_{\rm osc}(T_{\rm osc}) \left(\frac{T_{\rm grw}}{T_{\rm osc}}\right)^6\left(\frac{T_{\rm nr}}{T_{\rm grw}}\right)^4 \left(\frac{T_f}{T_{\rm nr}}\right)^3\nonumber\\
    &\simeq \rho_{\rm osc}(T_{\rm nr}) 
    \left( \frac{T_{\rm osc}}{T_{\rm nr}} \right)^3
    \left(\frac{T_{\rm grw}}{T_{\rm osc}}\right)^6\left(\frac{T_{\rm nr}}{T_{\rm grw}}\right)^4 \left(\frac{T_f}{T_{\rm nr}}\right)^3 \nonumber \\
    &\simeq \rho_{\rm osc}(T_f) 
    \,,
    \label{eq:rhofluc0}
\end{align}
where we have used $T_{\rm frag}\simeq \sqrt{\tilde \kappa_*} T_{\rm osc}$, Eq.~\eqref{eq:Tnr}, and the relation $\rho_{\rm fl} (T_{\rm grw})=\rho_{\theta} (T_{\rm grw}) = \rho_{\rm osc}(T_{\rm osc}) \left(\frac{T_{\rm grw}}{T_{\rm osc}} \right)^6$. 
Thus, we arrive at the following relation for the correction factor 
\begin{align}
    \mathcal{Z} \simeq 1 .
\end{align}
From this, we conclude that the modification factor in the DM relic density is order one, and we thus neglect its effect on our final parameter space. Note that this conclusion is in agreement with Ref. \cite{Eroncel:2022vjg}, where the modification factor to the DM relic density was also found to be $\mathcal{O}(1)$. To demonstrate this almost unchanged relic abundance, in Fig. \ref{Fig:rlcschm}, we show a schematic of the energy evolution of the pNGB with and without fragmentation (red and black lines, respectively). At $T \simeq T_{\rm grw}\simeq T_{\rm frag}$, the energy density is transferred to the fluctuations, which scale relativistically until $T_{\rm nr}$, after which they scale non-relativistically. As we saw from our calculation above, at $T_{\rm nr}$, the energy density of the fluctuations turns out to be the same as that of the background without fragmentation, thus following the background after $T_{\rm nr}$ and leading to unchanged relic abundance. The precise calculation of the modification factor involves non-linear lattice simulation, which we leave for future exploration.


\providecommand{\href}[2]{#2}\begingroup\raggedright\endgroup

\end{document}